\documentclass[12pt,letterpaper]{article}
\usepackage{amsmath,amssymb,array,calc,rotating,epsfig,psfrag, amscd}
\usepackage{color}
\usepackage[left=2cm,top=2cm,right=2cm,nohead]{geometry}
\usepackage[square, comma, compress]{natbib2}

\newcommand{\nc}{\newcommand}

\nc{\ra}{\rightarrow} 
\nc{\lra}{\leftrightarrow} 
\nc{\Ra}{\Rightarrow} 
\nc{\LRa}{\Leftightarrow} 
\nc{\blp}{{\big (}}
\nc{\brp}{{\big )}}
\nc{\Blp}{{\Big (}}
\nc{\Brp}{{\Big )}}
\nc{\bglp}{{\bigg (}}
\nc{\bgrp}{{\bigg )}}
\nc{\Bglp}{{\Bigg (}}
\nc{\Bgrp}{{\Bigg )}}
\nc{\slb}{{\rm [}}
\nc{\srb}{{\rm ]}}
\nc{\bslb}{{\rm \big [}}
\nc{\bsrb}{{\rm \big ]}}
\nc{\Bslb}{{\rm \Big [}}
\nc{\Bsrb}{{\rm \Big ]}}
\def\al{\alpha}

\def\eps{\epsilon}
\nc{\veps}{\varepsilon}
\def\gam{\gamma}

\def\lam{\lambda}
\def\om{\omega}
\nc{\vphi}{\varphi}
\def\tha{\theta}
\def\sig{\sigma}

\def\Gam{\Gamma}
\def\Lam{\Lambda}
\def\Om{\Omega}
\def\Sig{\Sigma}

\nc{\myvspace}{\rule[-1em]{0pt}{2.5em}}
\nc{\bea}{\begin{eqnarray}}
\nc{\eea}{\end{eqnarray}}
\nc{\be}{\begin{equation}}
\nc{\ee}{\end{equation}}
\nc{\barr}{\begin{array}}
\nc{\earr}{\end{array}}
\nc{\cA}{{\cal A}}
\nc{\cB}{ \cal B}

\def\cD{{\cal D}}

\nc{\cF}{{\cal F}}
\nc{\cG}{{\cal G}}

\nc{\cL}{{\cal L}}
\nc{\cM}{{\cal M}}
\def\N{{\cal N}}
\def\cN{{\cal N}}

\nc{\cQ}{{\cal Q}}
\nc{\cR}{{\cal R}}
\def\cS{{\cal S}}

\def\cU{{\cal U}}
\def\cV{{\cal V}}
\def\cV{{\cal V}}

\def\cZ{{\cal Z}}
\nc{\cQd}{\cQ^{\dagger}}
\nc{\cRd}{\cR^{\dagger}}
\nc{\BB}{{\mathbb B}}
\nc{\CC}{{\mathbb C}}
\nc{\DD}{{\mathbb D}}
\nc{\EE}{{\mathbb E}}
\nc{\FF}{{\mathbb F}}
\nc{\GG}{{\mathbb G}}
\nc{\HH}{{\mathbb H}}
\nc{\JJ}{{\mathbb J}}
\nc{\RR}{{\mathbb R}}
\nc{\PP}{{\mathbb P}}
\nc{\QQ}{{\mathbb Q}}
\nc{\ZZ}{{\mathbb Z}}
\nc{\calone}{{\mathbb 1}}

\nc{\half}{\frac{1}{2}}
\nc{\qrt}{\frac{1}{4}}
\nc{\del}{\partial}

\nc{\delbar}{\bar\partial}
\nc{\thalf}{\frac{t}{2}}
\nc{\Spin}{\operatorname{Spin}}
\nc{\SO}{\operatorname{SO}}

\nc{\Sp}{{\rm Sp}}
\nc{\com}[2]{{ \left[ #1, #2 \right] }}
\nc{\acom}[2]{{ \left\{ #1, #2 \right\} }}
\nc{\rr}{\rightarrow}
\nc{\p}{\partial}
\nc{\LT}{{\LL_\T}}
\nc{\Tr}{{\rm Tr}}
\nc{\tr}{{\rm tr}}
\def\com#1#2{{ \left[ #1, #2 \right] }}
\def\acom#1#2{{ \left\{ #1, #2 \right\} }}
\nc{\Adag}{A^{\dagger}}
\nc{\AdagI}{A^{\dagger I}}
\nc{\AdagJ}{A^{\dagger J}}
\nc{\AdagK}{A^{\dagger K}}
\nc{\AdagL}{A^{\dagger L}}
\nc{\AdagM}{A^{\dagger M}}
\nc{\Bdag}{B^{\dagger}}
\nc{\BdagI}{B^{\dagger}_I}
\nc{\BdagJ}{B^{\dagger}_J}
\nc{\BdagK}{B^{\dagger}_K}
\nc{\BdagL}{B^{\dagger}_L}
\nc{\BdagM}{B^{\dagger}_M}
\nc{\Cdag}{C^{\dagger}}
\nc{\CdagI}{C^{\dagger I}}
\nc{\CdagJ}{C^{\dagger J}}
\nc{\CdagK}{C^{\dagger K}}
\nc{\Ddag}{D^{\dagger}}
\nc{\DdagI}{D^{\dagger I}}
\nc{\DdagJ}{D^{\dagger J}}
\nc{\DdagK}{D^{\dagger K}}
\nc{\ttha}{\tilde{\theta}}
\nc{\tphi}{\tilde{\phi}}
\nc{\tsig}{\tilde{\sig}}
\nc{\tom}{\tilde{\om}}
\nc{\tlam}{\tilde{\lam}}
\nc{\tSig}{\widetilde{\Sig}}
\nc{\tPhi}{\tilde{\Phi}}
\nc{\tPhibar}{\ol{\tPhi}}
\nc{\tPi}{\tilde{\Pi}}
\nc{\tpsi}{\tilde{\psi}}
\nc{\tPsi}{\tilde{\Psi}}
\nc{\tgam}{\tilde{\gam}}
\nc{\tGam}{\tilde{\Gam}}
\nc{\tzeta}{\tilde{\zeta}}
\nc{\tZeta}{\tilde{\Zeta}}
\nc{\teta}{\tilde{\eta}}
\nc{\teps}{\tilde{\eps}}
\nc{\tEta}{\tilde{\Eta}}
\nc{\tchi}{\tilde{\chi}}
\nc{\tChi}{\tilde{\Chi}}
\nc{\Xit}{\tilde{\Xi}}
\nc{\tb}{\tilde b}
\nc{\tc}{\tilde c}
\nc{\te}{\tilde e}
\nc{\tf}{\tilde f}
\nc{\tg}{\tilde g}
\nc{\tj}{\tilde j}
\nc{\tp}{\widetilde{p}}
\nc{\tq}{\widetilde{q}}
\nc{\ts}{{\tilde s}}
\nc{\tu}{{\tilde u}}
\nc{\tv}{{\tilde v}}
\nc{\tw}{{\tilde w}}
\nc{\tx}{{\tilde x}}
\nc{\ty}{{\tilde y}}
\nc{\tz}{\tilde z}
\nc{\tA}{{\widetilde A}}
\nc{\tAbar}{{\ol \tA}}
\nc{\tB}{{\widetilde B}}
\nc{\tC}{{\widetilde C}}
\nc{\tD}{{\widetilde D}}
\nc{\tE}{{\widetilde E}}
\nc{\tG}{{\widetilde G}}
\nc{\tH}{{\widetilde H}}
\nc{\tJ}{{\widetilde J}}
\nc{\tJbar}{{\ol {\tilde J}}}
\nc{\tK}{{\widetilde K}}
\nc{\tL}{{\widetilde L}}
\nc{\tM}{{\widetilde M}}
\nc{\tN}{{\widetilde N}}
\nc{\tP}{{\widetilde P}}
\nc{\tQ}{{\widetilde Q}}
\nc{\tR}{{\widetilde R}}
\nc{\tS}{\widetilde{S}}
\nc{\tF}{\tilde{{\cal F}}}
\nc{\tX}{\widetilde{X}}
\nc{\tY}{\widetilde{Y}}
\nc{\tcZ}{\tilde{\cZ}}
\nc{\tcZbar}{\ol{\tcZ}}
\nc{\hb}{\hat b}
\nc{\hc}{\hat c}
\nc{\hd}{\hat d}
\nc{\he}{\hat e}
\nc{\hf}{\hat f}
\nc{\hg}{\hat g}
\nc{\hh}{\hat h}
\nc{\hp}{\hat p}
\nc{\hs}{\hat s}
\nc{\hv}{\hat v}
\nc{\hw}{\hat w}
\nc{\hx}{\hat x}
\nc{\hy}{\hat y}
\nc{\hz}{\hat z}
\nc{\zhat}{\hat z}
\nc{\hA}{\widehat{A}}
\nc{\hE}{\widehat{E}}
\nc{\hF}{\widehat{F}}
\nc{\hH}{\widehat{H}}
\nc{\hJ}{\widehat{J}}
\nc{\hK}{\widehat{K}}
\nc{\hL}{\widehat{L}}
\nc{\hM}{\widehat M}
\nc{\hN}{\widehat{N}}
\nc{\hV}{\widehat V}
\nc{\hcV}{\widehat \cV}
\nc{\hX}{\widehat X}
\nc{\hY}{\widehat Y}
\nc{\hZ}{\widehat Z}

\nc{\ha}{\widehat \alpha}
\nc{\hphi}{\hat{\phi}}
\nc{\hpsi}{\hat{\psi}}
\nc{\hgam}{\hat{\gam}}
\nc{\hPhi}{\hat{\Phi}}
\nc{\hPsi}{\hat{\Psi}}
\nc{\hGam}{\hat{\Gam}}
\nc{\omhat}{\hat{\om}}
\nc{\hOm}{\widehat{\Om}}

\nc{\w}{\wedge}

\nc{\vb}{\vec b}
\nc{\vc}{\vec c}
\nc{\vd}{\vec d}
\nc{\ve}{\vec e}
\nc{\vf}{\vec f}
\nc{\vg}{\vec g}
\nc{\vh}{\vec h}
\nc{\vp}{\vec p}
\nc{\vq}{\vec q}
\nc{\vr}{\vec r}
\nc{\vs}{\vec s}
\nc{\vv}{\vec v}
\nc{\vw}{\vec w}
\nc{\vx}{\vec x}
\nc{\vy}{\vec y}
\nc{\vz}{\vec z}

\nc{\vB}{\vec B}
\nc{\vC}{\vec C}
\nc{\vD}{\vec D}
\nc{\vE}{\vec E}
\nc{\vF}{\vec F}
\nc{\vG}{\vec G}
\nc{\vH}{\vec H}
\nc{\vP}{\vec P}
\nc{\vQ}{\vec Q}
\nc{\vR}{\vec R}
\nc{\vS}{\vec S}
\nc{\vV}{\vec V}
\nc{\vW}{\vec W}
\nc{\vX}{\vec X}
\nc{\vY}{\vec Y}
\nc{\vZ}{\vec Z}

\nc{\ol}{\overline}
\nc{\abar}{\ol{a}}
\nc{\bbar}{\ol{b}}
\nc{\cbar}{\ol{c}}
\nc{\dbar}{\ol{d}}
\nc{\ebar}{\ol{e}}
\nc{\fbar}{\ol{f}}
\nc{\ibar}{\ol{\imath}}
\nc{\jbar}{\ol{\jmath}}
\nc{\kbar}{\ol{k}}
\nc{\lbar}{\ol{l}}
\nc{\mbar}{\ol{m}}
\nc{\nbar}{\ol{n}}
\nc{\pbar}{\ol{p}}
\nc{\qbar}{\ol{q}}
\nc{\ubar}{\ol{u}}
\nc{\vbar}{\ol{v}}
\nc{\wbar}{\ol{w}}
\nc{\xbar}{\ol{x}}
\nc{\ybar}{\ol{y}}
\nc{\zbar}{\ol{z}}

\nc{\Abar}{\ol{A}}
\nc{\Bbar}{\ol{B}}
\nc{\Cbar}{\ol{C}}
\nc{\Dbar}{\ol{D}}
\nc{\Ebar}{\ol{E}}
\nc{\Fbar}{\ol{F}}
\nc{\Jbar}{\ol{J}}
\nc{\Kbar}{\ol{K}}
\nc{\Lbar}{\ol{L}}
\nc{\Mbar}{\ol{M}}
\nc{\Nbar}{\ol{N}}
\nc{\Pbar}{\ol{P}}
\nc{\Qbar}{\ol{Q}}
\nc{\Rbar}{\ol{R}}
\nc{\Sbar}{\ol{S}}
\nc{\Tbar}{\ol{T}}
\nc{\Ubar}{\ol{U}}
\nc{\Vbar}{\ol{V}}
\nc{\Wbar}{\ol{W}}
\nc{\Xbar}{{\overline X}}
\nc{\Ybar}{{\overline Y}}
\nc{\Zbar}{{\overline Z}}
\nc{\cZbar}{{\overline \cZ}}

\nc{\epsbar}{\ol{\epsilon}}
\nc{\lambar}{\ol{\lambda}}
\nc{\zetabar}{\ol{\zeta}}
\nc{\Zetabar}{\ol{\Zeta}}
\nc{\psibar}{\ol{\psi}}
\nc{\Psibar}{\ol{\Psi}}
\nc{\phibar}{\ol{\phi}}
\nc{\Phibar}{\ol{\Phi}}
\nc{\chibar}{\ol{\chi}}
\nc{\mubar}{\ol{\mu}}
\nc{\nubar}{\ol{\nu}}
\nc{\rhobar}{\ol{\rho}}
\nc{\ombar}{\ol{\om}}
\nc{\Ombar}{\ol{\Om}}
\nc{\Deltabar}{\ol{\Delta}}
\nc{\Thetabar}{\ol{\Theta}}
\nc{\xibar}{\ol{\xi}}
\nc{\Xibar}{\ol{\Xi}}

\nc{\Dthbar}{\ol{\rm D3}}

\nc{\gdot}{\dot{g}}
\nc{\xdot}{\dot{x}}
\nc{\ydot}{\dot{y}}
\nc{\phidot}{\dot{\phi}}
\nc{\sinp}{s_{\phi}}
\nc{\cosp}{c_{\phi}}
\nc{\tanp}{t_{\phi}}
\nc{\spone}{s_{\phi_1}}
\nc{\cpone}{c_{\phi_1}}
\nc{\tpone}{t_{\phi_1}}
\nc{\sptwo}{s_{\phi_2}}
\nc{\cptwo}{c_{\phi_2}}
\nc{\tptwo}{t_{\phi_2}}
\nc{\spth}{s_{\phi_3}}
\nc{\cpth}{c_{\phi_3}}
\nc{\tpth}{t_{\phi_3}}
\nc{\calp}{c_{\al}}
\nc{\salp}{s_{\al}}

\nc{\csch}{{\rm csch}}
\nc{\sech}{{\rm sech}}

\nc{\cothzlami}{\coth(z-\lam_i)}
\nc{\coshzlami}{\cosh(z-\lam_i)}
\nc{\sinhzlami}{\sinh(z-\lam_i)}

\nc{\cothzlamj}{\coth(z-\lam_j)}
\nc{\coshzlamj}{\cosh(z-\lam_j)}
\nc{\sinhzlamj}{\sinh(z-\lam_j)}

\nc{\cothlamij}{\coth(\lam_i-\lam_j)}
\nc{\coshlamij}{\cosh(\lam_i-\lam_j)}
\nc{\sinhlamij}{\sinh(\lam_i-\lam_j)}
\nc{\bah}{{\mathbf {\hat{A}}}}
\nc{\bX}{{\mathbf X}}
\nc{\ba}{{\bf a}}
\nc{\bb}{{\bf b}}
\nc{\bc}{{\bf c}}
\nc{\bd}{{\bf d}}
\nc{\bg}{{\bf g}}
\nc{\bk}{{\bf k}}
\nc{\bl}{{\bf l}}
\nc{\bm}{{\bf m}}
\nc{\bn}{{\bf n}}
\nc{\bo}{{\bf o}}
\nc{\bp}{{\bf p}}
\nc{\bq}{{\bf q}}
\nc{\br}{{\bf r}}
\nc{\bs}{{\bf s}}
\nc{\bt}{{\bf t}}
\nc{\bu}{{\bf u}}
\nc{\bv}{{\bf v}}
\nc{\bw}{{\bf w}}
\nc{\bx}{{\bf x}}
\nc{\by}{{\bf y}}
\nc{\bz}{{\bf z}}
\nc{\bom}{{\bf \om}}
\nc{\bombar}{{\mathbf \ombar}}
\nc{\bPhi}{{\bf \Phi}}

\nc{\rma}{{\rm a}}
\nc{\rmb}{{\rm b}}
\nc{\rmc}{{\rm c}}
\nc{\rmd}{{\rm d}}
\nc{\rmg}{{\rm g}}
\nc{\rk}{{\rm k}}
\nc{\rml}{{\rm l}}
\nc{\rmm}{{\rm m}}
\nc{\rmn}{{\rm n}}
\nc{\rmo}{{\rm o}}
\nc{\rmp}{{\rm p}}
\nc{\rmq}{{\rm q}}
\nc{\rmr}{{\rm r}}
\nc{\rms}{{\rm s}}
\nc{\rmt}{{\rm t}}
\nc{\rmu}{{\rm u}}
\nc{\rmv}{{\rm v}}
\nc{\rmw}{{\rm w}}
\nc{\rmx}{{\rm x}}
\nc{\rmy}{{\rm y}}
\nc{\rmz}{{\rm z}}
\nc{\Ffour}{{F^{(4)}}}
\nc{\Ffive}{{F^{(5)}}}

\nc{\dal}{\dot{\al}}
\nc{\thadot}{\dot{\tha}}
\nc{\thab}{\bar{\theta}}
\nc{\thal}{\theta^{\al}}
\nc{\thdal}{\bar{\theta}^{\dal}}

\nc{\thsigthm}{\tha \sigma^m \thab}
\nc{\thsigthn}{\tha \sigma^n \thab}

\nc{\Dal}{D_{\al}}
\nc{\Ddal}{\bar{D}_{\dal}}
\nc{\CDal}{{\cal D}_{\al}}
\nc{\CDdal}{\bar{\cal D}_{\dal}}

\nc{\eq}[1]{(\ref{#1})}
\nc{\non}{\nonumber}

\nc{\equ}{{\rm eq}}
\def\Im{{\rm Im ~}}
\def\ImN{{\rm Im \, \cN }}

\def\Re{{\rm Re ~}}

\nc{\vol}{{\rm vol}}
\nc{\Ainf}{A_{\infty}}
\nc{\End}{{\rm End}}
\nc{\Ext}{{\rm Ext}}
\nc{\IIB}{{\rm IIB}}
\nc{\Ad}{{\rm Ad}}
\nc{\IIA}{{\rm IIA}}
\nc{\AdS}{{\rm AdS}}
\nc{\CFT}{{\rm CFT}}
\nc{\Lifz}{{\rm Lif}_4(z)}
\nc{\Liftwo}{{\rm Lif}_4(2)}
\nc{\Schr}{{\rm Schr}_4(z)}
\nc{\Dslash}{\ensuremath \raisebox{0.025cm}{\slash}\hspace{-0.32cm} D}
\nc{\cDslash}{\ensuremath \raisebox{0.025cm}{\slash}\hspace{-0.32cm} \cD}
\nc{\no}{\!:\!\!}
\nc{\ointdz}{\oint\frac{dz}{2\pi i}}
\nc{\ointdzone}{\oint\frac{dz_1}{2\pi i}}
\nc{\ointdztwo}{\oint\frac{dz_2}{2\pi i}}
\nc{\ointdzb}{\oint\frac{d\zbar}{2\pi i}}
\nc{\ointdzbone}{\oint\frac{d\zbar_1}{2\pi i}}
\nc{\ointdzbtwo}{\oint\frac{d\zbar_2}{2\pi i}}
\nc{\dz}{\frac{dz}{2\pi i}}
\nc{\dzb}{\frac{d\zbar}{2\pi i}}
\nc{\bpm}{\begin{pmatrix}}
\nc{\epm}{\end{pmatrix}}
 \nc{\bitem}{\begin{itemize}}
 \nc{\eitem}{\end{itemize}}

\definecolor{cardinal}{rgb}{0.6,0,0}
\definecolor{darkgreen}{rgb}{0,0.5,0}
\definecolor{golden}{rgb}{0.92, 0.7, 0}
\definecolor{midnight}{rgb}{0, 0, 0.5}
\definecolor{darkblue}{rgb}{0.2, 0, 0.8}

\begin{document}
\begin{center}
\vskip 2 cm

{\Large \bf  Non-Relativistic Solutions of \\ \vskip 0.25cm 
$\N=2$ Gauged Supergravity }
\vskip 1.25 cm 
{Nick Halmagyi$^{a}$, Michela Petrini$^{a,b}$ and Alberto Zaffaroni$^{c}$}\\
\vskip 10mm

$^a$Laboratoire de Physique Th\' eorique et
Hautes Energies, \\   Universit\' e
Pierre et Marie Curie, \\ 
4 Place Jussieu, 75252 Paris Cedex 05,
France \\
\vskip 5mm
$^b$Institut de Physique Th\'eorique, \\
CEA Saclay, CNRS URA 2306, \\
F-91191 Gif-sur-Yvette, France \\
\vskip 5mm
$^c$Dipartimento di Fisica, Universit\`a di Milano--Bicocca, I-20126 Milano, Italy\\
            and\\
            INFN, sezione di Milano--Bicocca,
            I-20126 Milano, Italy
\vskip 5mm
halmagyi,\ petrini@lpthe.jussieu.fr, \\
alberto.zaffaroni@mib.infn.it \\

\end{center}

\begin{abstract}
We find infinite families of supersymmetric solutions of four dimensional, $\N=2$ gauged supergravity
with Lifshitz,  Schr\"odinger and also AdS symmetries. We focus on the canonical example of a single hypermultiplet and a single vector multiplet and find that the spectrum of solutions depends crucially on whether the gaugings are electric or magnetic but to a far milder extent on the strength of the gaugings. For purely electric or purely magnetic gaugings we generically find Lifshitz solutions, while for a mixed gauging we find Schr\"odinger and AdS solutions. 
For some of the gaugings  the theory has a known  lift to string/M-theory thus giving a higher dimensional embedding of our solutions. 
\end{abstract}
\section{Introduction}

The AdS/CFT duality \cite{Maldacena:1997re} relates quantum gravity on AdS spacetime to a relativistic quantum field theory on the boundary of AdS. While the canonical example of such  duality is between type IIB string theory on AdS$_5\times S^5$ and four dimensional $\N=4$ SYM, there are numerous generalizations  in various dimensions with less symmetry and a richer spectrum. 
Recently, it has been proposed to apply holography to problems in non-relativistic quantum field theory with applications to condensed matter physics.  
Gravitational duals  of non -relativistic field theories were first proposed \cite{Kachru:2008yh, Son:2008ye,Balasubramanian:2008dm,Koroteev:2007yp} by studying Einstein theory coupled to massive vectors\footnote{The original proposal of \cite{Kachru:2008yh} involved 
coupling gravity to a two-form  potential and a vector field, but in four dimensions the tensor can be dualized to a scalar.},  however it is of some interest to embed such solutions in a UV finite theory such as string theory or M-theory.  There has been significant progress made in this direction \cite{Herzog:2008wg, Maldacena:2008wh, Adams:2008wt, Hartnoll:2008rs, Taylor:2008tg, Donos:2009en, Gauntlett:2009zw,Bobev2009a, Bobev2009b,  Donos:2009xc, Donos:2009zf, Donos:2010tu, Donos:2010ax, Jeong:2009aa, O'Colgain:2009yd, Ooguri:2009cv, Colgain:2009wm, Balasubramanian:2010uk, Kraus:2011pf, Blaback:2010pp,Gregory:2010gx,Bobev:2011qx, Cassani:2011sv} and, indeed, this is the motivation for our current work. 
We will however take a somewhat different approach and study non-relavistic solutions  of $\N=2$ gauged supergravity in four dimensions, which should be holographically dual to the vacua of three-dimensional, non-relativistic quantum field theories. We will study gravitational duals to two kinds of non-relativistic solution, those with Lifshitz scaling \cite{Kachru:2008yh}
\be
t\ra \lam^z t\,,\ \ \ x^i\ra \lam x^i
\ee
combined with spatial rotations and those with Schr\"odinger scaling \cite{Son:2008ye,Balasubramanian:2008dm}
\be
x_+\ra \lam^z x_+ \,,\ \ x_-\ra \lam^{2-z}x_-\,,\ \  x^i\ra \lam x^i
\ee
combined with Galilean boosts. 

It has been well established that, for many internal manifolds, ten or eleven-dimensional supergravity can be consistently truncated to an effective gauged supergravity in lower dimensions. The prototypical examples are the maximal gauged supergravities in four, five and seven dimensions \cite{deWit:1982ig, Gunaydin:1984qu, Pilch:1984xy}, which are believed to be consistent truncations of IIB or eleven-dimensional supergravity on the appropriate dimensional sphere. For seven dimensional gauged supergravity the consistency of this reduction has been proved \cite{Nastase:1999cb}. These truncations keep all the lightest fields of just a subset of the various Kaluza-Klein towers, and,  when further truncated to a more manageable sector, have been used to extract precise results on holographic renormalization group flows \cite{Girardello:1998pd, Freedman:1999gp}. It should be noted that while the explicit lift of any given solution to these gauged supergravities to the higher dimensional theory is theoretically possible, it can be quite technically challenging and in some cases perhaps prohibitively so.

Another avenue by which one can construct gauged supergravity theories from string/M-theory is to use a set of fundamental forms on the internal manifold which possess a closed set of differential and algebraic relations \cite{Buchel:2006gb, Gauntlett:2009zw, Donos:2010ax, House:2005yc, KashaniPoor:2007tr, Cassani:2009ck}. This results in a truncation which is not so much a restriction to the lightest modes but instead to a singlet sector under a certain symmetry group. The invariant sector of a coset reduction is a prime example,  but more general geometric structures such as Sasaki-Einstein spaces and nearly Kahler manifolds also lead to such reductions.  If this set of fundamental forms is chosen judiciously, the spectrum is finite and typically these truncations include massive scalars and vector fields. For these consistent truncations, the lift to the higher dimensional theory is trivial since by construction the mapping is provided.

Thanks to these impressive works on consistent truncations, we have the confidence to explore the parameter space of gauged supergravity in general and, then,  to separately address the question of which subset of parameters admits an embedding into string/M-theory. We are not aware of a proof that any particular gauged supergravity cannot be embedded into string/M-theory, nor do we have any reason to suspect this may be true. It should be noted that while consistent truncations are useful for computing supersymmetric solutions, for non-supersymmetric solutions however, one must be wary of instabilities which lie outside the consistent truncation \cite{Bobev:2010ib} and accounting for these can re-introduce many of the complexities which had been  truncated away. 

In this paper, we discuss the conditions for supersymmetry for Lifshitz and  Schr\"odinger vacua in a general ${\cal N}=2$ gauged supergravity with vectors and hypermultiplets and we then apply the formalism to a specific example.
We will work with the canonical example of four dimensional  $\N=2$ gauged supergravity, namely we will retain just a single vector multiplet and a single hypermultiplet.  The virtue of this approach is that by fixing the field content of the gauged supergravity but allowing for a quite general gauging, we can simultaneously scan infinite families of string/M-theory compactifications\footnote{Some previous works in gauged supergravity where the charges have been left arbitrary are \cite{Behrndt:2000ph, Ceresole:2001wi, Hristov:2010eu}}. We find that while the conditions for supersymmetry produce vacua which are isolated in field space, they place extremely mild constraints on the charges and in this sense the vacua we find are somewhat universal.

Specifically, our scalar moduli space is
\be
{\cal M}_{SK} \times {\cal M}_Q = \frac{SU(1,1)}{U(1)}\times \frac{SU(2,1)}{SU(2)\times U(1)} \, , 
\ee
and we have two vector fields, the graviphoton and one from the vector multiplet.  We gauge two commuting isometries of the hypermultiplet moduli space allowing for both electric and magnetic charges, in the sense discussed in Section two. The type of vacua we find in any given theory depends crucially on whether the scalars are gauged electrically or magnetically, but depends only very mildly on the strength of the gaugings. 

 We can  gauge both  compact and non compact isometries of the hypermultiplet scalar manifold. We have found no interesting  supersymmetric non-relativistic solutions in the case with 
two non-compact gaugings. We  discuss in details the two cases of a pair of compact gauging, or one compact and one non-compact.
Interestingly  our results are essentially identical in both cases, indicating there is most likely a symmetry principle at work. Specifically, when all gaugings are either electric or magnetic, we find supersymmetric Lifshitz solutions with scaling parameter $z=2$ and we find that there are no supersymmetric Schr\"odinger or  $\mathcal{N}=2$ AdS solutions. These results indicate possible obstacles to constructing supersymmetric holographic RG-flows (along the line of  \cite{Girardello:1998pd, Freedman:1999gp})  between AdS and Lifshitz spacetimes\footnote{Such flows were found in a non-supersymmetric effective theory in \cite{Gubser:2009cg}.}.

On the other hand when the gaugings are a mixture of electric and magnetic, we find Schr\"odinger solutions and $\N=2$ AdS solutions but no Lifshitz solutions. We show that under mild conditions,  we can associate a supersymmetric Schr\"odinger solution to each ${\cal N}=2$ AdS$_4$  vacuum. The value of $z$ in the Schr\"odinger  solution is related to the mass ($m$) of the massive vector field in the corresponding AdS$_4$ vacuum by $z(z+1) = (mR)^2$,   as in the original construction in \cite{Balasubramanian:2008dm}.

For a few particular gaugings, these solutions have in fact been found before and in those cases we find precise agreement. In particular we find the $SU(3)\times U(1)$ invariant $\N=2$ AdS$_4$ vacuum of \cite{Warner:1983vz} and we also reproduce the Schr\"odinger solution found in \cite{Gauntlett:2009zw}. The Lifshitz solutions with $z=2$
found in \cite{Balasubramanian:2010uk, Donos:2010tu} are also  probably related  to our  class of solutions. Most if not all possible gaugings can be viewed as arising from a consistent truncation of string/M-theory, and, in those cases, by the very nature of consistent truncations, any solutions of our gauged supergravity can be claimed to be solutions of string/M-theory. But we leave a detailed analysis of this issue for further work.

This paper is organized as follows. In section two, we review some standard facts about four dimensional,  $\N=2$ gauged supergravity, largely to establish notation. In section three we study Lifshitz solutions, deriving and solving the conditions for supersymmetry. In section four we repeat this analysis for Schr\"odinger solutions and $\N=2$ AdS solutions. In section five we discuss the lift of these gauged supergravity theories to string/M-theory.

\section{Four Dimensional $\N=2$ Gauged Supergravity}

In the rest of the paper we will work in the framework  $\N=2$  gauged supergravity in four dimensions. We refer to \cite{Andrianopoli:1996vr, Andrianopoli:1996cm} for a detailed description of the formalism.

The fields of $\N=2$ supergravity are arranged into one graviton multiplet, $n_v$ vector multiplets
and $n_h$ hyper-multiplets.  The graviton multiplet contains the metric, the graviphoton, $A_\mu^0 $  and an $SU(2)$ doublet of gravitinos of opposite chirality, ($ \psi_\mu^A, \psi_{\mu \, A} $),
where  $A=1,2$ is an $SU(2)$ index.  
The vector multiplets consist of a vector, $A^I_\mu,$, two spin 1/2 of opposite chirality, transforming as an $SU(2)$ doublet, ($\lambda^{i \,A}, \lambda^{\bar{i}}_A$), 
and one complex scalar $z^i$. $A=1,2$ is the $SU(2)$ index, while $I$ and $i$ run on the number
of vector multiplets $I= 1, \dots, n_{\rm V}$,  $i= 1, \dots, n_{\rm V}$.   The scalar fields $z^i$ parametrise a special K\"ahler manifold of  complex dimension $n_{\rm V}$, $\mathcal{M}_{\rm SK}$. 
 Finally the hypermultiplets contain  two spin 1/2 fermions of opposite chirality,  ($\zeta_\alpha, \zeta^\alpha$),  and four real scalar fields, $q_u$, 
 where $\alpha = 1, \dots 2 n_{\rm H}$ and $u = 1, \ldots, 4 n_{\rm H}$.  
The scalars $q_u$ parametrise a quaternionic manifold of real dimension $4 n_{\rm H}$, $\mathcal{M}_{\rm Q}$. 

\vspace{0.4cm}

While in the  ungauged $\N=2$ supergravity the vector-  and the hyper-multiplets are decoupled at the two-derivative level, in the  gauged theory they have
non trivial interactions as can be seen from the bosonic Lagrangian
\bea
\label{boslag}
\mathcal{L}_{\rm bos} & =&  -\frac{1}{2} R + i ( \bar{\cN}_{\Lambda \Sigma } \cF^{- \Lambda}_{\,\, \mu \nu} \cF^{- \Sigma \mu \nu}  -   \cN_{\Lambda \Sigma} \cF^{+ \Lambda}_{\,\, \mu \nu} \cF^{+ \Sigma \mu \nu} ) \non \\
&& + g_{i \bar{j}} \nabla^\mu z^i \nabla_\mu \bar{z}^{\bar{j}} + h_{u v} \nabla^\mu q^u \nabla_\mu q^{v}   - \mathcal{V}(z, \bar{z}, q)   \, ,
\eea
where $\Lambda, \Sigma = 0, 1, \ldots, n_{\rm V}$.  The gauge field strengths are defined as
\be
\cF^{\pm \Lambda}_{\mu\nu} = \half \blp F^{\Lambda}_{\mu\nu}\pm \frac{i}{2}\eps_{\mu\nu \rho \sig}F^{\Lambda \rho\sig} \brp,
\ee
with $F^\Lambda_{\mu \nu} = \frac{1}{2} (\partial_\mu A^\Lambda_\nu - \partial_\nu A^\Lambda_\mu)$.  In this notation, $A^0$ is
 the graviphoton 
 and $A^\Lambda$, with $\Lambda = 1, \ldots, n_{\rm V}$, denote  the vectors in the vector multiplets. 
The period  matrix $\cN_{\Lambda \Sigma}$ is a function of the vector multiplet scalars. 

 $g_{i \bar{j}}$ and $h_{uv}$ are the metrics on the scalar manifolds $\mathcal{M}_{{\rm SK}}$ and $\mathcal{M}_{{\rm Q}}$, respectively.  
The covariant derivatives  are defined as 
\bea
\label{scalarder}
&& \nabla_\mu z^i = \partial_\mu z^i + k^i_{\, \, \Lambda} A^{\Lambda}_{\, \, \mu} \, ,  \\
&& \nabla_\mu q^u = \partial_\mu q^u + k^u_{\, \, \Lambda} A^{\Lambda}_{\, \, \mu} \, ,
\eea
where $k^i_\Lambda$ and $k^u_\Lambda$ are the Killing vectors associated to the isometries of the vector and hypermultiplet scalar manifold, respectively, that have been gauged.

In general, the gauge group can be at most a $(1 + n_{\rm V})$-dimensional subgroup $G$ of the isometry group of the scalar manifold
$\mathcal{M}_{{\rm SK}} \times \mathcal{M}_{{\rm Q}}$.
If the subgroup $G$ is non-abelian, it must necessary involve gaugings of the isometries of the vector multiplet space. In this paper we will work with abelian gaugings, so the vector multiplets are neutral and $G$ can be  identified with $1+ n_{\rm V}$ isometries of the
quaternionic manifold.  The Killing vector fields which generate these isometries admit a prepotential called the Killing prepotential. This is a set of real functions $P^x_\Lam$, where $x=1,2,3$ is an adjoint $SU(2)$ index, satisfying
\be
\Om^x_{uv}k^u_\Lam =-\nabla_v P^x_{\Lam} \, , 
\ee
where $\Om^x_{uv}$ and  $\nabla_v$ are the curvature and covariant derivative on $\cM_{{\rm Q}}$ (see appendix \ref{app:hyper} for more details on the Killing prepotentials). 

\vspace{0.2cm}

The scalar potential couples the hyper and vector multiplets,  and is given by
\be
\mathcal{V}(z, \bar{z}, q) = ( g_{i \bar{j}} k^i_\Lambda k^{\bar j}_\Sigma  
+ 4 h_{u v} k^u_\Lambda k^v_\Sigma ) \bar{L}^\Lambda L^\Sigma  + ( f_i^\Lambda g^{i \bar{j}} f^ \Sigma_{\bar j} - 3  \bar{L}^\Lambda L^\Sigma )
\mathcal{P}^x_\Lambda \mathcal{P}^x_\Sigma \, ,
\ee  
where $L^\Lambda$ are the symplectic sections\footnote{In the paper we will use both the symplectic sections 
$(L^\Lambda, M_\Lambda$) and the holomorphic sections
\be
(X^\Lambda, F_\Lambda) = e^{-K/2} (L^\Lambda, M_\Lambda) \, .
\ee}
 on $\mathcal{M}_{\rm SK}$ and  $f_i^\Lambda=   (\partial_i + \frac{1}{2}  \partial_i K)  L^\Lambda$, where $K$ is the vector multiplet K\"ahler potential.
 
\vspace{0.4cm}

The full Lagrangian is invariant under $\cN =2$ supersymmetry, with  supersymmetry variations for the fermionic fields given by
\bea
\label{gravitinoeq}
\delta \psi_{\mu A}&=& \mathcal{D}_\mu \epsilon_A + i   S_{AB} \gamma_\mu \epsilon^B  +  2i( \ImN)_{\Lambda \Sigma} L^{\Sigma} \cF_{\mu\nu}^{- \Lambda} \gamma^\nu \eps_{AB}  \eps^B \, , \\
\label{gauginoeq}
\delta \lam^{iA}&=& i \nabla_\mu z^i \gam^\mu \eps^A -g^{i\jbar} \fbar^{\Sigma}_{\jbar} \blp\ImN\brp_{\Sig \Lam} \cF^{- \Lambda}_{\mu\nu} \gamma^{\mu\nu }\eps^{AB}\eps_B  + W^{i A B} \epsilon_B\, , \\
\label{hyperinoeq}
\delta \zeta_{\al}&=& i\, \cU^{B\beta}_u\nabla_\mu q^u \, \gamma^\mu \epsilon^A \epsilon_{AB}\epsilon_{\alpha\beta} + N^{A}_{\al} \eps_A  \, , 
\eea
where $ \cU^{B\beta}_u$ are the vielbeine on the quaternionic manifold and 
\bea
S_{AB}&=&{{\rm i}\over2} (\sigma_x)_A^{\phantom{A}C} \epsilon_{BC}
{\cal P}^x_{\Lambda}L^\Lambda  \, , \nonumber\\
W^{iAB}&=&\epsilon^{AB}\,k_{\Lambda}^i \bar L^\Lambda\,+\,
{\rm i}(\sigma_x)_{C}^{\phantom{C}B} \epsilon^{CA} {\cal P}^x_{\Lambda}
g^{ij^\star} {\bar f}_{j^\star}^{\Lambda}    \, ,    \label{pesamatrice}\\
{\cal N}^A_{\alpha}&=& 2 \,{\cal U}_{\alpha u}^A \,k^u_{\Lambda} \, 
\bar L^{\Lambda} \, .
\non
\eea
In particular the covariant derivative on the spinors contains a contribution from the gauge fields
\be
\mathcal{D}_\mu \epsilon_A =  D_{\mu}\eps_A  + \frac{i}{2}  (\sig^x)_A^{\ B} A^\Lam_{\mu} P^x_{\Lam} \eps_B \, .
\ee

\vspace{0,4cm}

In some cases which we analyse later, the supersymmetry variations are not sufficient to fully determine the solutions, so we also list here the equations of motions.
The Einstein equation is 
\be
R_{\mu \nu} - \frac{1}{2} g_{\mu \nu} R =  T_{\mu \nu} \, , 
\ee
where the  energy momentum tensor is given by
\bea
T_{\mu \nu} &=& -  g_{\mu \nu}  \Big[ g_{i \bar{j} } \nabla^\rho z^i \nabla_\rho  \bar{z}^{\bar j} +  h_{u v} \nabla^\rho q^u \nabla_\rho q^{v} +  \ImN_{\Lambda \Sigma } F^\Lambda_{\,\, \rho \sig} 
F^{\Sigma \rho \sig}  -\mathcal{V}(z, \bar{z}, q)  \Big] \non \\
&& + 2 \Big[ g_{i \bar{j} } \nabla^\mu z^i \nabla_\nu \bar{z}^{\bar j} +  h_{u v} \nabla_\mu q^u \nabla_\nu q^{v}  + 2  \, \ImN_{\Lambda \Sigma } F^\Lambda_{\,\, \mu \rho} F_\nu^{\Sigma \rho} \Big] \, .
\eea
The equations of motion for the gauge fields are
\be
\partial_\mu (\sqrt{- g} \, \Im G_\Lambda^{- \mu \nu} ) = - \frac{\sqrt{- g}}{2}   (g_{i \bar{j}}  k^i_{\,\, \Lambda}  k^{\bar j}_{\,\, \Sigma}  A^{\Sigma \nu} + 
 h_{u v}  k^u_{\,\, \Lambda}  k^{v}_{\,\, \Sigma}   A^{\Sigma \nu} ) \, ,
\ee
with $G_\Lambda^{- \mu \nu}  = \bar{\mathcal{N}}_{\Lambda \Sigma} F^{\Lambda  \mu \nu}$.

\subsection{The Canonical Model}

Having set up the general machinery, we now specialize to a particularly simple example of $\cN =2$ gauged supergravity in four dimensions, namely that of one vector multiplet ($ n_{\rm V} =1$) and one hypermultiplet
($ n_{\rm H} =1$). In spite of its simplicity, we will see that it exhibits a rich spectrum of supersymmetric solutions.  In this case we take the scalar manifold of the theory to be 
\be
\label{scalarman}
{\cal M}_{{\rm SK}} \times {\cal M}_{{\rm Q}} = \frac{SU(1,1)}{U(1)}\times \frac{SU(2,1)}{SU(2)\times U(1)} \, .
\ee

The  vector multiplet sector contains a single complex scalar and a natural choice of coordinates on $\mathcal{M}_{\rm SK}$ is   $z = \tau$, where $\tau$ parametrizes the upper-half plane.
With this choice, the K\"ahler potential and the metric are
\bea
\label{kpot}
&& K=- 3\log \bslb i(\tau-\ol{\tau}) \bsrb\, , \\
 \label{vecmet} 
&& {\rm d} s^2 = \frac{3}{4}\frac{{\rm d} \tau  {\rm d} \bar{\tau}}{(\Im \tau)^2}  \, .
\eea 

The four scalars of the hypermultiplet parametrise  ${\cal M}_{\rm Q}$.  There are several alternative way to choose coordinates on  ${\cal M}_{\rm Q}$.  In this paper, depending on the model we study, we will consider the following possibilities
\begin{itemize}
\item[{\it a})]  two complex coordinates
\be
\label{complchyp}
\{q^u \} \quad \leftrightarrow \quad (\zeta_1, \zeta_2 ) \, .
\ee
With this parametrisation the metric on ${\cal M}_Q$ becomes
\bea
\label{cchypermet}
{\rm d}s^2 = \frac{{\rm d}\zeta_1 {\rm d}\zetabar_1+ {\rm d}\zeta_2 {\rm d}\zetabar_2}{1-|\zeta_2|^2 - |\zeta_2|^2}
+\frac{(\zeta_1 {\rm d} \zetabar_1+\zeta_2 d\zetabar_2)(\zetabar_1 {\rm d} \zeta_1+\zetabar_2 {\rm d} \zeta_2)}{(1-|\zeta_2|^2 - |\zeta_2|^2)^2}.
\eea

\item[{\it b})] one complex and two real coordinates\footnote{These two co-ordinate systems are related by
\be
\xi=\frac{\zeta_2}{1+\zeta_1}\,, \quad \quad
\rho = \frac{1-|\zeta_1|^2-|\zeta_2|^2}{|1+\zeta_1|^2}\,, \quad \quad
\sig = \frac{i(\zeta_1-\zetabar_1)}{|1+\zeta_1|^2}\non \,.
\ee}
\be
\label{gchyp}
\{q^u \} \quad \leftrightarrow \quad (\xi, \rho, \sigma ) \, .
\ee
Then the metric on ${\cal M}_Q$  is 
\be
\label{gchypermet}
{\rm d}s^2 = \frac{1}{4 \rho^2}  ({\rm d}\rho)^2 +  \frac{1}{4 \rho^2} [ {\rm d} \sigma - i (\xi {\rm d} \bar{\xi} -\bar{\xi} {\rm d} \xi) ]^2 + \frac{1}{\rho} {\rm d} \xi {\rm d} \bar{\xi}  \, .
\ee
\end{itemize}

The $\cN=2$ theory is completely determined only after the gaugings and the
sections  $(X^\Lambda, F_\Lambda$)  have been fixed.   These may or may not  be compatible with a prepotential $\cF(X^\Lam)$ \cite{Ferrara:1995xi}, in that $F_{\Lam}=\del_{X_\Lambda} \cF$. 
Typically, Kaluza-Klein reductions of string or M-theory, lead to four dimensional effective actions with a prepotential  that is a cubic function of the $X^\Lambda$, 
and both electric and magnetic gaugings. Indeed it is by now well established that internal fluxes can generate magnetic 
gaugings in the lower dimensional theory. One can keep both electric and magnetic gaugings  by considering   
Lagrangians where the hypermultiplet scalars corresponding to the symmetries that are magnetically realised
 are dualised into tensors (see for instance \cite{Sommovigo:2004vj, Dall'Agata:2003yr, deWit:2005ub}).  
However, it is always possible to transform a generic dyonic gauging into a purely electric one, by a symplectic transformation on the sections ($X^\Lambda, F_\Lambda$)
 \be
 (X^\Lambda, F_\Lambda) \quad \mapsto \quad   (\tilde{X}^\Lambda, \tilde{F}_\Lambda) =  \cS (X^\Lambda, F_\Lambda) \, ,
 \ee
 where the matrix 
 \be
 \cS =  \bpm A& B \\C & D \epm
 \ee
 is an element of $Sp( 2 + 2 n_{\rm V}, \mathbb{R})$. This transformation leaves the K\"ahler potential invariant, but changes the period matrix
 $\cN_{\Lambda  \Sigma}$ by a fractional transformation
 \be
 \cN_{\Lambda \Sigma} (X, F)  \quad \mapsto  \quad   \tilde{\cN}_{\Lambda \Sigma} (\tilde{X}, \tilde{F}) = ( C + D   \cN_{\Lambda \Sigma} (X, F)) (A + B  \cN_{\Lambda  \Sigma} (X, F))^{-1} \, .
 \ee
 
Our strategy will then be to consider purely electric gaugings, 
allowing for sections $(\tX^{\Lam},\widetilde{F}_{\Lam})$ which are a general symplectic rotation of those
obtained from the cubic prepotential. More precisely we start by choosing a cubic prepotential
\be
\cF= -\frac{X_1^3}{X_0}\, ,
\ee
with sections $(\Lambda = 0,1)$
\be
\label{el}
\begin{array}{l}
X^{\Lam}= (1,\tau)\, , \\
F_{\Lam}=(\tau^3, -3\tau^2 )\, .
\end{array} 
\ee
The period matrix\footnote{When the holomorphic sections are compatible with the existence of a holomorphic  prepotential ${\cal F}(X)$ such that
$F_\Lambda = \partial_\Lambda {\cal F}(X)$, the gauge kinetic matrix $\cN_{\Lambda \Sigma}$ can be written in terms of derivatives of the prepotential
\be
\cN_{\Lambda \Sigma} = \bar{F}_{\Lambda \Sigma }  +2 i  \frac
{{\rm Im} F_{\Lambda {\rm M}}  \, {\rm Im}F_{\Sigma  \Upsilon} X^{{\rm M}} X^{\Upsilon}}
 {{\rm Im }F_{{\rm M} \Upsilon} X^{\rm M} X^{\Upsilon}}  \, .
\ee
where $F_{\Lambda \Sigma} =  \partial_{\Lambda \Sigma}^2 {\cal F}$.}
in this case takes the form
\be
\label{Nel}
\cN_{\Lambda \Sigma} =  \frac{1}{2} \bpm - \tau^3 -  3 \tau^2 \bar{\tau} & 3 \tau (\tau + \bar{\tau}) \\
3 \tau (\tau + \bar{\tau}) & - 3 ( 3 \tau + \bar{\tau})  \epm \, .
\ee

This choice corresponds to electric gaugings for both the graviphoton and the vector $A^1$.  A dyonic configuration where the graviphoton is electrically gauged and $A^1$ magnetically
will correspond to a symplectic rotation with 
\be
\label{gravel}
\cS_1 = \bpm  1 &0 & 0 & 0 \\
0 & 0 & 0 & 1\\
0 & 0 & 1 & 0 \\
0 & -1 & 0 & 0 \epm  \, , 
\ee
while the converse case, with a magnetic graviphoton and electric $A^1$, is obtained by setting 
\be
\label{gravmag}
\cS _2= \bpm  0 &0 & 1 & 0 \\
0 & 1 & 0 & 0 \\
-1 & 0 & 0 & 0 \\
0 & 0 & 0 & 1  \epm \, .
\ee
Finally, purely magnetic gaugings correspond to the rotation
\be
\label{mag}
\cS _3= \bpm  0 &0 & 1  & 0 \\
0 & 0 & 0 & 1  \\
-1 & 0 & 0 & 0 \\
0 & -1 & 0 & 0 \epm \, .
\ee

We will see  that these four different cases lead to very different patterns of solutions.  The first and fourth cases, \eqref{el} and \eqref{mag}, allow for families of $\Lifz$ vacua, while 
the second and third, \eqref{gravel} and \eqref{gravmag}, give families of AdS$_4$ and $\Schr$ vacua.

\vspace{0.4cm}

As already mentioned, we only consider abelian gaugings of the hypermultiplet isometries.  
From \eqref{scalarman}, it is easy to see that the isometry group
of the quaternionic manifold is $SU(2,1)$. The full set of corresponding Killing vectors and prepotentials are presented in appendix \ref{app:hyper}. In fact we will just utilize the following three  Killing vectors
\be
\label{compactkv}
\begin{array}{l}
k_3= - \frac{i}{2} ( - \zeta_1 \del_{\zeta_1} + \zeta_2 \del_{\zeta_2}-c.c. ) \,,  \\
k_4= - \frac{i}{2} ( \zeta_1 \del_{\zeta_1} + \zeta_2 \del_{\zeta_2}-c.c. )\, ,   \\
k_6=\frac{i}{2} [ (1+\zeta_1^2)\del_{\zeta_1} + \zeta_1 \zeta_2\del_{\zeta_2}-c.c. ] \, .
\end{array} 
\ee
The vector fields $(k_3,k_4)$ generate compact isometries while $k_6$ is a non-compact generator.

Since the theory contains two vectors, the graviphoton and the vector in the vector multiplet, we can gauge at most two isometries. We will consider in detail the case of two compact gaugings, and also the case of one compact and one non-compact gauging:
\begin{enumerate}
\item  {\bf Two compact gaugings}  \\
\vskip -3mm
This case is best described choosing complex coordinates, \eqref{complchyp}, for the hypermultiplet. We choose the two $U(1)$'s associated with the Killing vectors $k_3$ and $k_4$, which
correspond to rotations of the phases of the coordinates $\zeta_1$ and $\zeta_2$. The two  gaugings  are defined as 
\be
k_\Lambda = a_\Lambda \frac{( -k_3+k_4 )}{2} + b_\Lambda \frac{(k_3+k_4)}{2}  \, ,
\ee
where $a_\Lambda$ and $b_\Lambda$ are the electric (magnetic) charges.
The corresponding  Killing prepotentials are
\be
{\cal P}^x_\Lambda = a_\Lambda P^x_a + b_\Lambda P^x_b \, ,
\ee
with 
\bea
P_{a}&=&\frac{1}{(|\zeta_1|^2+|\zeta_2|^2)\sqrt{1-|\zeta_1|^2-|\zeta_2|^2}} 
\bpm 
-\Im (\zeta_1 \zeta_2)  \\ 
 \Re  (\zeta_1 \zeta_2)  \\
\frac{-|\zeta_2|^4+|\zeta_2|^2-|\zeta_1|^2(1+|\zeta_2|^2)}{2\sqrt{1-|\zeta_1|^2-|\zeta_2|^2}} 
\epm\,, \\
P_{b}&=&\frac{1}{(|\zeta_1|^2+|\zeta_2|^2)\sqrt{1-|\zeta_1|^2-|\zeta_2|^2}} 
\bpm 
\Im (\zeta_1 \zeta_2)  \\ 
-\Re  (\zeta_1 \zeta_2)  \\ 
\frac{-|\zeta_1|^4+|\zeta_1|^2-|\zeta_2|^2(1+|\zeta_1|^2)}{2\sqrt{1-|\zeta_1|^2-|\zeta_2|^2}} 
\epm  \,.
\eea

\item {\bf One  compact and one non compact  gauging} \\
\vskip -3mm
 For this choice, the coordinates \eqref{gchyp} are more suitable. The compact isometry corresponds to the sum of the two commuting compact generators $k_3$ and $k_4$ 
\be
k_\xi = i( \xi\del_{\xi}-\xibar\del_{\xibar}) =-(k_3+k_4) 
\ee
while the non compact isometry corresponds to shifts of the coordinate $\sigma$
\be
\label{sigmashift}
k_\sigma = \del_\sig =- \frac{1}{2} k_3 + \frac{3}{2} k_4 -k_6 \, .
\ee
The corresponding Killing prepotentials are given by
\be
P_\sigma = \bpm 0 \\ 0 \\- \frac{1}{2 \rho} \epm \,  \qquad \qquad 
P_\xi =  \bpm \frac{\xi + \bar{\xi}}{\sqrt{\rho}}  \\  \frac{ \bar{\xi} - \xi}{i \sqrt{\rho}} \\   \frac{| \xi|^2}{ \rho} - 1  \epm \, .
 \ee
Then we define the generic gauging as
\be
k_\Lambda = a_\Lambda k_\sigma + b_\Lambda k_\xi \, , \qquad \Lambda = 0,1 \, ,
\ee
with Killing prepotential
\be
{\cal P}^x_\Lambda = a_\Lambda P^x_\sigma + b_\Lambda P^x_\xi \, .
\ee

\end{enumerate}

As already mentioned in the Introduction, one could a priori also consider gauging two non compact isometries of the hypermultiplet
manifold, \eqref{scalarman}. These can be chosen to be the shift of the coordinate $\sigma$ as defined in \eqref{sigmashift} and 
\be
k_2= \frac{1}{2} [\del_\xi + \del_{\bar{\xi}} - i (\xi - \bar{\xi})  \del_\sigma  ] \, ,
\ee
corresponding to the sum of the Killing vectors $k_2$ and $k_7$ in \eqref{Kv}. Such gaugings appear naturally in some
dimensional reduction of type IIA theory on coset and Nearly K\"ahler manifolds  and provide examples of ${\cal N}=1$ AdS vacua \cite{KashaniPoor:2007tr,Cassani:2009ck,Cassani:2009na}. However, we have found neither interesting
supersymmetric  Lifshitz and Schr\"odinger  solutions nor ${\cal N}=2$ AdS vacua in the case of two non compact gaugings.

\section{ Supersymmetric Lifshitz Solutions} \label{sec:Lifshitz}

In this section we compute supersymmetric $\Lifz$ solutions of $\N=2$ gauged supergravity with one hyper-and one vector multiplet. But before reducing to this simple case, 
we consider some general features of Lifshitz solutions which hold for a generic number of multiplets.  A four-dimensional space-time  with Lifshitz symmetry of degree $z$   
\be (t\, , x\, , y\, , r\, ) \, \rightarrow (\lambda^z \, t\, , \lambda\, x \, , \lambda\, y \, , \lambda^{-1} \, r \, ) \, , \ee
is given by  \cite{Kachru:2008yh}   
 \be
{\rm d}s^2 = R^2\left (r^{2 z} {\rm d}t^2 - \frac{{\rm d}r^2}{r^2} - r^2 {\rm d}x^2 -r^2 {\rm d}y^2 \right ) \, .
\ee

In order to preserve the scaling symmetry,  all the scalar fields $z^{i}$ and $q^u$ must be constant 
\be z^i = z^i_0\, , \qquad\qquad q^u = q^u_0. \ee
 The interesting terms in the Lagrangian (\ref{boslag})  (i.e. setting all fermions and scalar derivatives to zero) are
 \be
\label{efflag}
-\frac{1}{2} \cR  +  {\rm Im} {\cal N}_{\Lambda \Sigma} F_{\mu\nu}^{\Lambda}F^{\mu\nu\, \Sigma} + h_{uv} k^u_\Lambda k^v_\Sigma A^\Lambda A^\Sigma  - \cV(z,\bar z, q) \, .
\ee
As discussed above we will consider exclusively the case of Abelian gaugings and thus only the scalars in the  hypermultiplets are charged. 

We  look for solutions  where the gauge fields have only temporal component \cite{Kachru:2008yh,Taylor:2008tg}
\be
\label{Alif}
A^\Lambda_t =  r^z A^\Lambda \, , 
\ee
so that the only non trivial component of the gauge field strength is 
\be
\label{Flif}
F^\Lambda_{r\, t } = \frac{z}{2} A^\Lambda r^{z-1} \, .
\ee
Einstein's equations are then algebraic
\bea
\label{E1}
&& \ImN_{\Lambda \Sigma } A^\Lambda A^\Sigma = - \frac{ (z-1)}{z} R^2 \, , \\
\label{E2}
&& h_{u v}  k^u_{\,\, \Lambda}  k^{v}_{\,\, \Sigma} A^\Lambda  \,  A^\Sigma  =  (z-1) \,, \\
\label{E3}
&& V = -\frac{z^2 + z + 4}{2\, R^2} \, ,
\eea
and Maxwell's equations reduce to 
\be
\label{max1}
h_{u v} k^u_\Lambda k^v_\Sigma A^\Sigma = - \frac{z}{R^2} \ImN_{\Lambda \Sigma} A^{\Sigma} \, . 
\ee
Notice that  not all these equations are independent. By contracting Maxwell's equations with $A^\Lambda$ we recover one component of Einstein's equation. Finally, the equations of motion for the scalar fields require 
\be 
\partial_{z^i} \cV_{eff} =0 \, , \qquad\qquad \partial_{q^u} \cV_{eff} =0 \, , 
\ee
where we have defined the effective potential
\be
\cV_{eff}(z, \bar z, q)  =  \cV(z, \bar z, q) - \frac{ z^2}{R^4} \ImN_{\Lambda \Sigma} A^\Lambda A^{\Sigma}+ \frac{2}{R^2} h_{u v} k^u_\Lambda k^v_\Sigma(\zeta)  A^\Lambda A^\Sigma\, . 
\ee

Due to the nature of our ansatz, we obtain  algebraic, not differential, equations
for certain real constants. The number of equations precisely matches  the number of unknowns: we have $n_V$ constants $A^\Lambda$
corresponding to the electric profile for each gauge field and 
$n_V$ constraints from Maxwell's equations. We have $n_s = 2 n_V + 4 n_H$ constants $z^i_0\, , q^u_0$
from each real scalar $z^i\, , q^u$ and $n_s$ constraints coming from the derivatives of $V_{eff}=0$. We have three constraints from the $R_{xx},R_{00},R_{rr}$ components of Einstein's equation
but one of these is implied by tracing over the Maxwell equations. Now since there are two constants in the gravity theory $(z,\Lambda)$, in total we have $n_V+n_s+2$ equations for the same number of constants.

\subsection{Conditions for Supersymmetric Lifshitz Solutions}
\subsubsection{The gravitino equation}

With the choice of   frames
\be e^0= R r^z {\rm d}t\, , \, e^1= R r {\rm d}x\, , \, e^2= R r {\rm d}y\, , \, e^3=R \frac{{\rm d}r}{r} \, , 
\ee 
the 0-,1- and 3-components of the gravitino equation \eqref{gravitinoeq} are\footnote{Recall that in our conventions the gauge field configurations entering the supersymmetry variations are given by (\ref{Alif}) and (\ref{Flif}),
so that  ${\cal F}_{rt}^{\Lambda\, -}= \frac{z}{4} A^\Lambda r^{z-1}$ and $ {\cal F}_{xy}^{\Lambda\, -}= -  \frac{i z}{4} A^\Lambda r^{2}$ .} 
\bea
&&\gamma^0 \partial_0 \epsilon_A +\frac{i}{2 R}  (\sigma^x)_{A}^{\phantom{A}B} A^\Lambda 
P_\Lambda^x \gamma^0 \epsilon_B  + \frac{z}{2 R} \gamma^3 \epsilon_A - \frac{i z}{2 R^2} \mathcal{N}
\epsilon_{A B} \gamma^{03} \epsilon^B + i  S_{A B} \epsilon^B =0 \, , \label{grav0} \\
&&\gamma^1 \partial_1 \epsilon_A  + \frac{1}{2 R} \gamma^3 \epsilon_A + \frac{i z}{2 R^2} \mathcal{N}
\epsilon_{A B} \gamma^{03} \epsilon^B + i  S_{A B} \epsilon^B =0 \, , \label{grav1} \\
&& \gamma^3 \partial_3 \epsilon_A - \frac{i z}{2 R^2} \mathcal{N}
\epsilon_{A B} \gamma^{03} \epsilon^B + i  S_{A B} \epsilon^B =0 \, , \label{grav3} 
\eea
where we have defined
\be
{\cal N}\equiv {\rm Im} {\cal  N}_{\Sigma\Lambda} L^\Lambda A^\Sigma\, .
\ee  

When $z=1$ we recover AdS$_4$-spacetime which requires a separate treatment (see section \ref{schrsection}), here will restrict to
the Lifshitz case $z> 1$. We choose a radial profile for the supersymmetry parameters\footnote{A phase in the spinor, 
$\epsilon_A\sim r^{\frac{a +i f }{2}} \epsilon_A^0$,  is forbidden by  the simultaneous presence of  $\epsilon^A$ and its  conjugate $\epsilon_A$ in the  supersymmetry conditions  and by the fact that the scalars have no radial profile. An $r$-dependent  phase in $\epsilon_A$ and in the phase $\theta$, that will be shortly introduced,
presumably plays a role in solutions describing renormalization group flows. We should note indeed the similarity of the $\Lifz$ supersymmetry conditions with analogous ones for black holes \cite{Dall'Agata:2010gj}.}
\be\epsilon_A = r^{\frac{a}{2}} \epsilon^0_A
\ee
where $\epsilon^0_A$ is a constant spinor.  Comparing  equations \eq{grav0} and \eq{grav3} we find
\be
\label{e1}
\frac{i}{2 R}  P_{\Lambda}^x A^{\Lambda}  (\sigma^x)_A^{\phantom{A}B}  \gamma^0 \epsilon_B =
 \frac{a-z}{2R} \gamma^3 \epsilon_A\, ,
\ee
which by compatibility immediately implies $z=a$ and the constraint
 \be
P_{\Lambda}^x A^{\Lambda}  =0 \, .
\ee
The remaining gravitino equations now give
\bea
\label{newpr}
\epsilon_A &=&  \frac{4 i  R}{1+z } S_{AB} \gamma^3 \epsilon^B \,,\\
\label{newpr2}
\epsilon_A &=& -  \frac{2 i  z }{R (z-1)} {\cal N} \gamma^0 \epsilon_{AB} \epsilon^B .
\eea

There are also some compatibility conditions associated to these projectors. First, by squaring  \eqref{newpr2} we find 
\bea
|\mathcal{N}|^2 =  \frac{(z-1)^2 R^2}{ 4 z^2} &&
\Rightarrow \ \  {\cal N} = e^{i \theta} \frac{(z  -1) R}{ 2 z} \, \label{gravit}.
\eea
The phase $e^{i\theta}$ can be reabsorbed with a K\"ahler transformation in the Lagrangian and we will set it to one in the following \footnote{Such K\"ahler transformation is equivalent to a redefinition of the sections $L^\Lambda$. In the search for the most general solution the sections in (\ref{el}) should be allowed to have an arbitrary overall phase $e^{i\theta}$.  However, we have found no interesting Lifschitz solutions with $\theta\ne 0$.}.

We thus have
\be
\epsilon_A =  - i   \epsilon_{AB} \gamma^0 \epsilon^B \, .
\ee 

The mutual  compatibility of the $\gamma^3$ and $\gamma^0$ projections gives
\be
\label{newpr3}
\epsilon_A = H_A^{\phantom{A}B} \gamma^{3 0} \epsilon_B \, ,
\ee
with 
\be
H_A^{\phantom{A}B} =  \frac{4  R  }{1 + z  }  S_{A C}\epsilon^{C B}  \equiv h^x (\sigma_x)_A^{\phantom{A}B} \, ,
\ee
where 
\be
h^x= \frac{ 2 R \, i  }{1+ z }  {\cal P}^x_{\Lambda} L^{\Lambda} \,.
\label{e2}
\ee
By squaring (\ref{newpr}) and (\ref{newpr3}) we obtain
\bea
 \frac{16 R^2}{(1+z)^2}  S_{AB}S^{BC} =  \delta_A^C\, , \qquad& \Longrightarrow& \qquad 
\sum_{i=1}^3 | h^x |^2 = 1 \, , \nonumber\\
 H_A^{\phantom{A}B} H_B^{\phantom{B}C} =  \delta_A^{\phantom{A}C}\, ,\qquad &\Longrightarrow& \qquad 
\sum_{x=1}^3 ( h^x   )^2  =1\, .
\eea
These conditions immediately imply $\sum_{i=1}^3 ({\rm Im} h^x)^2 =0$ and thus
\be
\Im h^x=0 \,  \label{Imhx} \, ,
 \ee
requiring  that  $h^x$ is a real three-vector of length one. Consequently $H_A^{\phantom{A}B}$ is an hermitian matrix with eigenvalues $\pm 1$. 

The conditions of supersymmetry are now fully compatible and can be reduced to the canonical form
 \bea
\tilde \epsilon_A &=&  (\sigma^3)_A^B  \gamma^0\gamma^3\tilde\epsilon_B \, , \nonumber\\
\tilde \epsilon_A & =&- i  \epsilon_{AB} \gamma^0 \tilde \epsilon^B \, , 
\label{spin}
\eea
by a unitary change of basis,  thus demonstrating that any solution would be $\frac{1}{4}$-BPS or, in
other words, preserves two real supercharges. $\bar \epsilon \gamma^\mu \epsilon$ gives the Killing vector $\partial/\partial_t$ as expected, in agreement with what  found in ten-dimensional solutions 
\cite{Donos:2010tu}.

\subsubsection{The gaugino and hyperino equations}

The gaugino equations give
\be
W^{i\, AB} \epsilon_B  +   \frac{z}{R^2}  {\cal N}^i \gamma^0\gamma^3 \epsilon^{AB} \epsilon_B =0 \, , 
\ee
where
\be
{\cal N}^i\equiv g^{i\, \bar j} \bar f_{\bar j}^\Sigma {\rm Im} {\cal N}_{\Sigma\Lambda} A^\Lambda \, , 
\ee
and the gamma matrices can be eliminated using the gravitino conditions to yield
\be
( -\frac{z }{ R^2 }  {\cal N}^i \epsilon^{AC} H_C^{\phantom{C}B} +  W^{i\, AB} ) \epsilon_B =0 \, . 
\ee

The hyperino variations give 
\be
R\, {\cal N}_\alpha^A \epsilon_A + i\, {\cal U}^{B\beta}_u k^u_\Lambda A^\Lambda \epsilon_{AB} \epsilon_{\alpha\beta} \gamma^0 \epsilon^A  = 0 \, ,
\ee
and, once more,  by eliminating the gamma matrices we obtain
\bea 
&& \left  ( \epsilon_{\alpha\beta}\, {\cal U}_u^{A\beta} k^u_\Lambda A^\Lambda  +  R\,  {\cal N}_\alpha^A  \right ) \epsilon_A =0\,.
\label{dilhypa}
\eea
Since the spinors $(\eps_1,\eps_2)$ are independent in a  $\frac{1}{4}$-BPS solution, the matrix expressions in brackets should vanish identically and the
hyperino equation becomes
\be 
k^u_\Lambda \left ( A^\Lambda + 2 R \, \bar L^\Lambda \right )= 0 \, .
\ee

\subsubsection{Some general properties of the supersymmetry conditions}

In total, the full set of conditions for supersymmetric Lifshitz solutions is
\bea
\sum_{x=1}^3 ( h^x   )^2  &=& 1\,, \\
\Im h^x&=&0\,, \\
P_{\Lambda}^x A^{\Lambda}  &=&0\,, \\
{\cal N} &=& \frac{(z  -1) R}{ 2 z} \, ,  \label{gravitb} \\
W^{i\, AB}&=& \frac{z }{ R^2 }  {\cal N}^i \epsilon^{AC} H_C^{\phantom{C}B}  \, ,\label{grav2} \\
k^u_\Lambda \left ( A^\Lambda + 2 R \, \bar L^\Lambda \right )&=& 0 \, \label{hyp}.
\eea

 A simple way to solve equation (\ref{hyp})   would be to set 
$A^\Lambda= -2 R  \bar L^\Lambda$. Using equation  (\ref{gravitb}), we obtain 
\be  
-  \frac{1}{2} {\rm Im} {\cal N}_{\Sigma\Lambda} A^\Lambda A^\Sigma = \frac{z-1}{2 z} R^2 \, , 
\ee
which correctly reproduces the equation of motion  (\ref{E1}).  However the condition 
\be
{\rm Im} {\cal N}_{\Sigma\Lambda} L^\Lambda \bar L^\Sigma\equiv -1/2 \, ,
\ee
which is valid for all ${\cal N}=2$ supergravities,  gives the unphysical value $z=-1$. We conclude that, in order to find interesting $\Lifz$ solutions,  we need to find loci
on the hypermultiplet manifold where the Killing vectors $k^u_\Lambda$ degenerate or become aligned.

In the following we deal with cases where, on  the relevant scalar locus,  ${\cal P}^x_\Lambda$ points in a particular direction in the $x$ space, say the $x=3$ direction.  Then $h^{1}=h^2=0$ and we need to require $h^3=1$. The full set of gravitino conditions become
\begin{equation}
 P^3_\Lambda A^\Lambda = 0\, ,  \qquad   2 i R \,  P^3_\Lambda L^\Lambda =  z+1\, , \qquad  \frac{2 z}{R} {\rm Im} {\cal  N}_{\Sigma\Lambda} L^\Lambda A^\Sigma = z-1  \, .
\label{GRAV}
\end{equation}
 We should also impose the gaugino and hyperino conditions (\ref{grav2}) and (\ref{hyp}) 
 \be 
 \label{simp} i   P_\Lambda^3 \bar f_{\bar j}^\Lambda +\frac{z}{R^2}  {\rm Im} {\cal N}_{\Sigma\Lambda} \bar  f_{\bar j}^\Lambda A^\Sigma =0 \, , \qquad k^u_\Lambda \left ( A^\Lambda + 2 R \, \bar L^\Lambda \right )= 0  \, ,
 \ee
 and the Maxwell equations (\ref{max1}).

\subsection{Lif$_4$(z) Vacua from Canonical Gaugings}\label{Lifsol}

We now restrict our analysis to the theory with only one vector and one hypermultiplet and  show that there is a  $\Lifz$ solution in the case of  a cubic prepotential with purely electric gaugings
 or the case of the symplectic rotation \eq{mag},  which is equivalent to purely magnetic gaugings. These solutions exist only  for $z=2$ but with  very mild constraints on the gauging parameters.

\subsubsection{Compact gaugings}

We first consider purely electric gaugings.
Consider first  the hyperino variation \eq{hyp}
 \bea
&& [ a_0 (A^0 + 2 R  \bar L^0) + a_1 (A^1 + 2 R \bar L^1) ] \zeta_1 =  0 \, , \\
&& [ b_0 (A^0 + 2 R \bar L^0) + b_1 (A^1 + 2 R \bar L^1) ]  \zeta_2= 0 \,.
\eea
As mentioned above, a solution with  $\zeta_1 \ne 0 $ and $\zeta_2\ne 0$ leads to unphysical values for $z$. Similarly the choice $\zeta_1=\zeta_2=0$ does not lead to a solution
since all quantities in the previous formulae  are real except for 
\be 
\label{secrot}
L(\tau) =e^{K/2}(1,\tau) \, , 
\ee 
and $\Im\tau=0$ is a singular point of the metric  \eq{vecmet}. We therefore conclude that either
\be
(\zeta_1, b_1)=(0,0) \ \ {\rm or}\ \ (\zeta_2, a_1)=(0,0).
\ee
These two choices are clearly symmetric and we choose the latter. On the locus $\zeta_2=0$ the only non zero component of the Killing prepotentials is ${\cal P}^3_\Lambda$. 

From the gravitino conditions (\ref{GRAV}),  the gaugino and hyperino conditions (\ref{simp}),  and Maxwell's equation (\ref{max1}) we find (with $b_1<0$)

\bea
A_0&=&  -\frac{R}{\sqrt{2} ({\rm Im \tau})^{3/2}} =  \frac{432}{R^2 \, b_1^3} \, , \\
A_1&=&   - \frac{ 2 \left ( 216 \, b_0 -108 \, a_0 \pm \sqrt{11664  \, a_0^2 + \,  R^4 \, b_1^6}\right )}{R^2\,  b_1^4} \, , \\
{\rm Im} \, \tau &=& \frac{R^2 \, b_1^2}{72} \, , \\
{\rm Re} \, \tau &=& \frac{A_1}{A_0}  \, , \\
|\zeta_1|^2&=& \frac{-a_0 \, A_0 + b_0 \, A_0 + b_1 \, A_1}{b_0 \, A_0 + b_1\,  A_1} \, , \\
\zeta_2&=&0\,.
\eea
The only constraint on the parameters comes from $0\le |\zeta_1|<1$,  which can be satisfied for a large choice of gauging parameters. We have checked that the second order equations of motion are all satisfied. The solution found in \cite{Balasubramanian:2010uk, Donos:2010tu} falls in this class of vacua, which as we now see exists quite generally for electric gaugings with canonical prepotential. 

Similar $\Lifz$ solutions also exist if we perform the simultaneous  symplectic rotation  \eq{mag} on $A^0$ and $A^1$, which is equivalent to a cubic prepotential with purely magnetic gaugings. The sections are now  
\be 
L(\tau) =e^{K/2}(\tau^3, - 3 \tau^2)
\ee and we still find a solution  for
\be
(\zeta_1, b_1)=(0,0) \ \ {\rm or}\ \ (\zeta_2, a_1)=(0,0) \, .
\ee
Choosing again the second option  we find
\bea
A_0&=& \frac{16}{R^2 \, b_1^3} \, , \\
A_1&=& - \frac{ 2 \sqrt{3}}{  b_1} \, , \\
{\rm Im} \, \tau &=&  \sqrt{3} {\rm Re} \, \tau =  \frac{6}{R^2 \, b_1^2} \, , \\
{\rm Re} \, \tau &=&  -\frac{3}{4} \frac{A_0}{A_1} = \frac{ 2 \sqrt{3}}{R^2 \, b_1^2} \, , \\
|\zeta_1|^2&=& \frac{-a_0 \, A_0 + b_0 \, A_0 + b_1 \, A_1}{b_0 \, A_0 + b_1\,  A_1} \, , \\
\zeta_2&=&0 \, , 
\eea
and in addition we have to impose an algebraic relation between the gaugings
\be a_0= \frac{32\,  b_0^2 - 8\,  \sqrt{3}\,  R^2\,  b_0 \, b_1^3 + R^4\,  b_1^6}{32\,  b_0 - 4 \, \sqrt{3}\,  R^2\,  b_1^3}\,. \ee 

With similar arguments one can check that there are no $\Lifz$ solutions with a single symplectic rotation on just $A^0$,  
\eq{gravmag}, or $A^1$, \eq{gravel}. In particular, since the $SU(3)$-invariant  sector of ${\cal N}=8$ gauged supergravity has one electric and one magnetic gauging \cite{Bobev:2010ib} this demonstrates that there are no $\Lifz$ solutions in this theory.

\subsubsection{One non-compact  gauging}

When one isometry is non compact we obtain almost identical results and thus we will be brief. The non trivial constraints from  
the hyperino variation are now \eq{hyp}
 \bea
&& [a_0 (A^0 + 2 R \bar L^0) + a_1 (A^1 + 2 R \bar L^1) ]  = 0 \ , \\
&& [b_0 (A^0 + 2 R \bar L^0) + b_1 (A^1 + 2 R \bar L^1) ] \xi= 0  \, .
\eea

In the case of electric gaugings with $L(\tau) =e^{K/2}(1,\tau)$ we can solve the hyperino equations with $\xi=0$ and $a_1$=0. The other equations then require
\bea
A_0&=&   -\frac{R}{\sqrt{2} ({\rm Im \tau})^{3/2}} = -\frac{54}{R^2 \, b_1^3} \, ,\\
A_1&=&   \frac{ 54  \, b_0}{R^2\,  b_1^4} - \frac{1}{b_1}  \, , \\
{\rm Im} \, \tau &=& \frac{R^2 \, b_1^2}{18} \, , \\
{\rm Re} \, \tau &=& \frac{A_1}{A_0} = -\frac{b_0}{b_1} + \frac{R^2\, b_1^2}{54} \, , \\
\rho &=& -\frac{ a_0\,  A_0}{2 ( b_0\, A_0 + b_1 \, A_1)} =  - \frac{27 a_0}{ R^2 b_1^3} \,  , \\
\xi&=&0\, .
\eea

The case of a double symplectic rotation can appear when studying type IIA solutions with Roman mass.
We have  $ L(\tau) =e^{K/2}(\tau^3, - 3 \tau^2)$, we are still forced to set $a_1=0$.
The solution is
\bea
A_0&=&   -\frac{2}{R^2 \, b_1^3} \, , \\
A_1&=&   \frac{\sqrt{3}}{b_1} \, , \\
{\rm Im} \, \tau &=& \sqrt{3} {\rm Re} \, \tau = \frac{3}{2\, R^2 b_1^2} \, ,\\
{\rm Re} \, \tau &=&  -\frac{3}{4} \frac{A_0}{A_1} =   \frac{ \sqrt{3}}{2\, R^2\, b_1^2} \, , \\
\rho &=& -\frac{ a_0\,  A_0}{2 ( b_0\, A_0 + b_1 \, A_1)} =  \frac{a_0}{ R^2 b_1^3} \, , \\
\xi&=&0 \, , 
\eea
with the constraint 
\be
b_0= \frac{\sqrt{3}-1}{2} R^2 \, b_1^3 \, .
\ee
Once again we have found  no $\Lifz$ solutions with a single symplectic rotation. In all cases we have checked that the second order equations of motion are satisfied.

\section{Supersymmetric  AdS$_4$ and Schr$_4(z)$ Solutions}
\label{schrsection}

We discuss now the case of AdS$_4$ and $\Schr$ solutions. We treat them simultaneously since in our formalism the supersymmetry conditions  for ${\cal N}=2$ AdS$_4$  and $\Schr$ are very similar.
We will not discuss  ${\cal N}=1$ AdS$_4$ solutions\footnote{Examples of this class of solutions in related contexts
can be found in \cite{ KashaniPoor:2007tr,Cassani:2009ck, Bobev:2010ib, Cassani:2009na, Bobev:2009ms}.
}, where the conditions for  supersymmetry typically require proportionality between $\epsilon^1$ and $\epsilon^2$. 

\subsection{The Schr$_4$  space-time}
We first recall the form of a $\Schr$ solution \cite{Son:2008ye, Balasubramanian:2008dm}
\be
{\rm d}s^2 = R^2 (r^{2 z} {\rm d}x_+^2 - 2 r^2 {\rm d}x_+ {\rm d}x_- -\frac{{\rm d}r^2}{r^2} - r^2 {\rm d}x^2  ) \, ,
\ee
which is invariant under the scaling symmetry
\be (x_+\, , x_-\, , x\, , r\, ) \, \rightarrow (\lambda^z \, x_+\, , \lambda^{2-z}\, x_- \, , \lambda\, x \, , \lambda^{-1} \, r \, ) \, . \ee
This is a solution of Einstein's equation with a cosmological constant and a massive vector. We again set  all the scalar fields $z^{i}\, , q^u$ to be constant 
and we deduce the equations of motion from the Lagrangian (\ref{efflag}). The gauge fields are now 
\be
A_{+}^\Lambda = A^\Lambda r^z \,.
\ee
Einstein's equation is
 \bea
 \label{ES1}
2 h_{u v}  k^u_{\,\, \Lambda}  k^{v}_{\,\, \Sigma} A^\Lambda   A^\Sigma - \frac{z^2}{R^2}  \ImN_{\Lambda \Sigma } A^\Lambda A^\Sigma &=& 2 z^2 - z -1  \, , \\
 \label{ES2}
V &=& - \frac{3}{R^2}  \, , 
\label{eqsSc1}
\eea
and Maxwell's equation is
\begin{equation}
\label{Smax}
2 h_{u v}  k^u_{\,\, \Lambda}  k^{v}_{\,\, \Sigma} A^\Sigma = - \frac{z(z+1)}{ R^2}  \ImN_{\Lambda \Sigma }  A^\Sigma\, .
\end{equation}
In contrast to the Lifshitz solutions, there is no contribution from the gauge fields to the potential since $F_{\mu\nu}F^{\mu\nu} = A_\mu A^\mu=0$  and thus we have the scalar equations of motion
\be \partial_{z^i} V =0 \, , \qquad\qquad \partial_{q^u} V =0\,. \ee

\subsection{Conditions for Supersymmetric  Schr$(z)$  Solutions}

It is convenient to  work with the null frames
\be 
e^+= \frac{1}{2} R r^z {\rm d}x_+\, ,\ \  e^-= R \left ( r^{z} {\rm d}x_+ - 2\, r^{2 -z} {\rm d}x_-\right )\, , \ \  e^2= R r {\rm d}y\, , \ \  e^3=R \frac{{\rm d}r}{r} \, ,
\ee 
so that the metric becomes
\be 
{\rm d}s^2 = 2 e^+ e^- - (e^2)^2 - (e^3)^2 \, .
\ee 

Chosing a spinor that satisfies
\be \gamma^+  \epsilon^A=  0 \, ,
\ee
the components of the gravitino equation \eqref{gravitinoeq} reduce to  
\bea
 && \partial_+ \epsilon_A +\frac{i}{ R}  (\sigma^x)_{A}^{\phantom{A}B} A^\Lambda 
P_\Lambda^x  \epsilon_B  + \frac{1}{2 R} \gamma^{-3} \epsilon_A   - \frac{i z\mathcal{N}}{ R^2} 
\left (  \gamma^{3}+  i \gamma^2\right ) \epsilon_{A B}\epsilon^B + i  S_{A B}\gamma_+  \epsilon^B =0 \,  ,\non\\
&& \partial_- \epsilon_A  = 0  \, , \non \\
&& \partial_2 \epsilon_A  - \frac{1}{2 R} \gamma^{23} \epsilon_A -  i  S_{A B}\gamma^2  \epsilon^B =0 \, , \non \\
&& \partial_3 \epsilon_A  + \frac{1-z}{2 R} \gamma^{-+} \epsilon_A  - i  S_{A B}\gamma^3 \epsilon^B =0 \, .\label{gravS} 
\eea
We can solve these conditions with 
\be
\epsilon_A = r^{\frac{2-z}{2}} \epsilon^0_A
\ee
and 
\bea
i S_{AB} \epsilon^B &=& - \frac{1}{2 R} \gamma ^3 \epsilon_A\,,\\
(\sigma^x)_{A}^{\phantom{A}  B} A^\Lambda P^x_\Lambda \epsilon_B &=& \frac{2  z {\cal N}}{R} \gamma^3 \epsilon_{AB} \epsilon^B.
\eea
Consistency of these equations leads to
\begin{equation}
\left (P^x_\Lambda A^\Lambda + 2 z {\cal N} P^x_\Lambda \bar L^\Lambda \right ) (\sigma^x)_{A}^{\phantom{A} B} \epsilon_B =0.
\end{equation}
Since $\gamma^+ \epsilon=0$, the gauge field contribution drops out  of the gaugino and hyperino equations
\bea 
&&    D^{i\, AB}  \epsilon_B =0\,,\\
&& {\cal N}_\alpha^A    \epsilon_A =0.
\label{dilhypb}
\eea

When looking for ${\cal N}=2$ AdS$_4$ and $\Schr$ vacua we will consider the  spinors $(\eps_1,\eps_2)$ 
as independent. The conditions for supersymmetry are then
\bea
\label{S1} i S_{AB} \epsilon^B &=& - \frac{1}{2 R} \gamma ^3 \epsilon_A \,, \\
 \label{S2} P^x_\Lambda A^\Lambda  &=&- 2  z {\cal N} P^x_\Lambda \bar L^\Lambda \, , \\  
\label{S3}  P^x_\Lambda \bar f^\Lambda_{\bar j} &=&0\, ,  \\
 \label{S4} k^u_\Lambda \bar L^\Lambda  &=& 0 \, ,
\eea
which should be supplemented by Maxwell's equation (\ref{Smax}). The spinor bilinear $\bar \epsilon \gamma^\mu \epsilon$ gives the Killing vector $\partial/\partial_-$,  associated with the number operator, as also found in ten-dimensional solutions 
\cite{Donos:2009zf}.

The AdS solutions have $z=1$ and $A^\Lambda=0$. Moreover the condition $\gamma^+\epsilon_A=0$ is superfluous
and we have four independent real spinors; to these Poincar\'e supersymmetries we need to add the four superconformal ones which depend explicitly on $(x,y,t)$. The AdS Killing spinors satisfy indeed $D_\mu \epsilon_A=\gamma_\mu \epsilon_A$. 
A class of ten dimensional Schr\"odinger backgrounds  with $z=2$ admit  additional  Poincar\'e and also superconformal symmetries \cite{Donos:2009zf}; it would be interesting to see if there is a similar phenomenon in  ${\cal N}\ge 2$ gauged supergravities.

\subsubsection{Relation between AdS$_4$ and Schr$_4$ vacua}

A close relation between AdS and $\Schr$ vacua  is expected \cite{Maldacena:2008wh, Donos:2009en,Bobev2009b} as has been recently discussed in great detail \cite{Kraus:2011pf}.
Here we will analyse it at the level of gauged supergravity, focusing on a theory with a single vector multiplet. 

Suppose that we start with an ${\cal N}=2$ AdS$_4$ vacuum satisfying (\ref{S1}),(\ref{S3}) and (\ref{S4}). 
These conditions do not depend explicitly on the vector fields $A^\Lambda$ and are identical for the ${\cal N}=2$ AdS$_4$ and Schr$(z)_4$ cases.  Therefore, we would expect that
for every ${\cal N}=2$ AdS$_4$ solution there exists  a corresponding 
supersymmetric $\Schr$ one, with the same radius $R$ and the same value for the scalar fields, provided that the Maxwell's equations  (\ref{Smax}), 
the gravitino constraint (\ref{S2}),
and the equations of motion can be satisfied for a choice of $A^\Lambda$.  
We now show that under mild conditions,  this is  the case.

Multiplying Maxwell's equations  (\ref{Smax}) by $L^\Lambda$ and using the hyperino condition (\ref{S4}) we find $z(z+1){\cal N}=0$. Excluding uninteresting solutions with $z=0$ or $z=-1$  we reduce the gravitino constraint  (\ref{S2}) to
\begin{equation}
\label{PP} 
P^x_\Lambda \, A^\Lambda \, = \, {\cal N}\, =\, 0\,.
\end{equation}
We see from equation  (\ref{S3}) that the $\bar f^\Lambda_{\bar i}$ have a common phase. Setting
\begin{equation}
\label{AA}
A^\Lambda = c f^\Lambda_i \, 
\end{equation} 
where $c$ is a complex constant to make $A^\Lambda$ real, we solve all equations in (\ref{PP}).  The first equation becomes equivalent to the gaugino condition (\ref{S3}) and the second one follows from the special geometry identity $ {\rm Im} {\cal N}_{\Lambda\Sigma} L^\Lambda f^\Sigma_i =0$ \cite{Andrianopoli:1996vr,Andrianopoli:1996cm}. 

Multiplying Maxwell's equations  (\ref{Smax}) by $A^\Lambda$  and using the identity  $ {\rm Im} {\cal N}_{\Lambda\Sigma} \bar f^\Lambda_{\bar j} f^\Sigma_i= -\frac{1}{2} g_{i\bar j}$ \cite{Andrianopoli:1996vr,Andrianopoli:1996cm} we find a quadratic equation for $z$
\begin{equation}
\label{zeta}
z^2 + z - 4 R^2  g^{i\bar i} h_{uv} k^u_\Lambda k^v_\Sigma  f^\Lambda_i  \bar f^\Sigma_{\bar i} =0 
\end{equation}
and so we  find a solution with positive $z$  whenever $k_u^\Lambda  f^\Lambda_{ i} $ is non vanishing. This fact has a 
simple interpretation. In the ${\cal N}=2$ AdS$_4$ vacuum we have a massless graviphoton and a massive vector with $m^2= 4  g^{i\bar i} h_{uv} k^u_\Lambda k^v_\Sigma  f^\Lambda_i  \bar f^\Sigma_{\bar i}$, as it can be easily checked by diagonalizing the kinetic term in (\ref{efflag}). The equation for $z$ can be then written as 
\be
z ( z+1 ) = (mR)^2\,.
\ee 
We see that the exponent $z$ is related to the mass of the vector fields in the corresponding AdS$_4$ vacuum,   as in the original construction in \cite{Balasubramanian:2008dm}.  Finally, the Einstein's equation (\ref{ES1}) will fix the normalization of $A^\Lambda$  ($z\ge1$ is required for consistency). 

This demonstrates that under mild conditions,  we can associate a supersymmetric Schr\"odinger solution to each ${\cal N}=2$ AdS$_4$  vacuum.  These results hold for a generic number of hypermultiplets.

\subsection{AdS$_4$ and Schr$_4$ Vacua in the Canonical Model}\label{AdSSchr}
From the Lagrangians which arise from consistent truncations of M-theory compactifications on Sasaki-Einstein manifolds \cite{Gauntlett:2009zw} one finds ${\cal N}=2$ AdS$_4$  solutions at the origin of moduli space. In our language, these gauged supergravities correspond 
to the case of a single symplectic rotation \eq{gravel} with a particular choices of charges. In this section we focus again on the theory with one vector and one hypermultiplet.  We show that in the case  of a single symplectic rotation, \eq{gravel} or \eq{gravmag},  there are ${\cal N}=2$ AdS$_4$ and $\Schr$ vacua for a large set of gauging parameters. We found no ${\cal N}=2$ AdS$_4$ or $\Schr$ vacua  in the cases of purely electric gaugings and  of a double symplectic rotation,  where we found $\Lifz$ solutions. One also finds other interesting $\N=2$ ${\rm AdS}_4$ vacua in the $SU(3)$ sector of the $\N=8$ theory \cite{Warner:1983vz,Bobev:2010ib} and here we show that these vacua also exist  for a very general set of gaugings.

We will discuss in details the case of a  symplectic rotated prepotential corresponding to an electric-magnetic duality on the vector $A^1$. The case where the graviphoton is rotated is completely analogous. 

\subsubsection{Compact gaugings}

It is still useful to start with the  hyperino equation which is just $k^u_\Lambda L^\Lambda=0$, or, explicitly, 
\bea
&& [a_0 L^0(\tau) + a_1 L^1(\tau) ] \zeta_1 =0   \, , \\
&& [b_0 L^0(\tau) + b_1 L^1(\tau) ]  \zeta_2 = 0   \, . 
\eea
Now we have $L=e^{K/2}( 1, -3 \tau^2)$, corresponding to an electric-magnetic duality on $A^1$.

We first consider ${\cal N}=2$ AdS$_4$ solutions and find two different such vacua. One is at the origin of the hypermultiplets  $\zeta_1=\zeta_2=0$ and the vector multiplet scalar $\tau$ is fixed by the gaugino variation to be
\be \tau = i \sqrt{\frac{a_0-b_0}{a_1-b_1}} \, . \ee
The gravitino equation simply sets the scale of $R$,
\be \frac{1}{R^2}= \frac{1}{2}\sqrt{(a_0-b_0)(a_1-b_1)^3}\, .\ee 

There is  another ${\cal N}=2$ vacuum away from the origin. If we set $\zeta_1=0$ we can still solve the hyperino conditions 
by choosing
\be 
\tau = i \sqrt{-\frac{b_0}{3 b_1}} \, .
\ee
The gaugino condition then fixes
\be 
|\zeta_2|^2= \frac{3 \, a_0\,  b_1 + a_1 \, b_0 - 4\,  b_0\,  b_1}{3\,  a_0 \, b_1 + a_1\,  b_0} \, , 
\label{z2}\ee
 and the gravitino equation simply sets the scale of $R$
 \be \frac{1}{R^2} = \frac{ 3 \sqrt{3} (a_1\, b_0 - a_0\,  b_1)^2}{32 \sqrt{- b_0^3 \, b_1}} \, .\ee
There is an equivalent solution with $\zeta_2=0$. 
 
An example of the model with one magnetic and one electric gauging is the $SU(3)$-invariant sector of  ${\cal N}=8$ gauged supergravity. The values of the gauging parameters in the $SU(3)$-invariant sector
can be determined by comparison with reference \cite{Bobev:2010ib}, where the action has been written as an ${\cal N}=2$ 
gauged supergravity. They are proportional to $(a_0,a_1)=(1,0)$ and $(b_0,b_1)=(1/2,-\sqrt{3}/2)$\footnote{To compare with the notations in reference \cite{Bobev:2010ib}  we need to perform a further (purely electric) rotation on the vectors $A^0$ and $A^1$.}. The  vacua that we found above have $(\zeta_1,\zeta_2)=(0,0)$  and  $(\zeta_1, \zeta_2)=(0,1/\sqrt{3})$
and the  ratio of the values of the potential in the two vacua is equal to $3\sqrt{3}/4$. 
These numbers  precisely correspond  to those for the ${\cal N}=8$ vacuum with $SO(8)$ global symmetry and the IR ${\cal N}=2$ solution with  $SU(3)\times U(1)$ global symmetry in the $SU(3)$-invariant  sector of  ${\cal N}=8$ gauged supergravity  \cite{Bobev:2010ib}.
 We see that the existence of a pair of ${\cal N}=2$  AdS$_4$  vacua is quite general and holds for almost arbitrary  values of the gaugings. 

We now consider $\Schr$ solutions. 
It is obvious from (\ref{zeta})   that the solution in the origin, with $(\zeta_1,\zeta_2)=(0,0)$, can only give solutions with $A^\Lambda \ne0$ in the unphysical case $z=0,z=-1$. Both vectors fields are in fact massless at the origin. On the other hand, in the case with $\zeta_1=0$ and $\zeta_2$ given in (\ref{z2}), we can find a solution; from (\ref{AA}) and (\ref{zeta}) we see that
\be 
A_0 = \frac{3 b_1}{b_0} A_1  
\ee
and that $z$ solves the algebraic equation
\be
z^2\, + z\, -\frac{4 ( 3 \, a_0\, b_1 + a_1\, b_0)( 3 \, a_0\, b_1 + a_1\, b_0- 4 b_0 b_1)}{3 (a_1\, b_0 - a_0\, b_1)^2}  =0 \, .
 \ee
 The equations of motion are satisfied and one of them fixes the value of $A^1$.  For a large choice of gauging parameters we can find physical solutions. We note that the charges corresponding to the $SU(3)$-invariant sector yield solutions with $z=(-2.56,1.56)$.

The case of a rotation of the graviphoton is similar and there are analogous solutions. In the case of a cubic prepotential with electric gaugings or the case of a double electric-magnetic rotation instead we found no interesting
solutions. 

\subsubsection{One compact and one non compact gauging}

The case where one of the isometry is non compact is almost identical and we will be brief. Again there are ${\cal N}=2$ AdS$_4$ and $\Schr$ solutions for  one electric and one magnetic gauging. We discuss as before  the case of an electric-magnetic duality on $A^1$. 

There is an ${\cal N}=2$ AdS$_4$ vacuum for $\xi=0$ and
\be \tau = i\sqrt{- \frac{a_0}{3 a_1}} \, , \qquad \rho= -\frac{2 a_0 a_1}{3 a_1 b_0 + a_0 b_1} \, , \ee
with radius
\be  
\frac{1}{R^2} = \frac{ 3 \sqrt{3} (a_1\, b_0 - a_0\,  b_1)^2}{8 \sqrt{- a_0^3 \, a_1}} \, .
\ee

For the same values of the scalar fields there is   a $\Schr$ solution with
\be A_0 = \frac{3 a_1}{a_0} A_1 \ee
and $z$ determined by
\be 
z^2\, + z\, -\frac{4( 3 \, a_1\, b_0 + a_0\, b_1)^2}{3 (a_1\, b_0 - a_0\, b_1)^2}  =0 \, .
\ee

The model with one symplectically rotated vector  appears in the Lagrangian corresponding to the consistent truncation  of M-theory compactified on a Sasaki-Einstein manifold $SE_7$ \cite{Gauntlett:2009zw}. The reduction naturally gives a cubic prepotential and a tensor field; the tensor field can be dualized to the scalar $\sigma$ with a simultaneous dualization of $A^1$. With our normalizations, the gaugings
are  proportional to $(a_0,a_1)=(6 \sqrt{2}, -2 \sqrt{2})$ and $(b_0,b_1)= (-\sqrt{2},0)$\footnote{Reference  \cite{Gauntlett:2009zw} uses a different symplectic rotation given in equation (2.38) of the same reference; the gauging parameters reported above have been correspondingly rotated with respect to those in \cite{Gauntlett:2009zw}.}.  The AdS vacuum has $\tau=i$ and $\rho=4$ as in \cite{Gauntlett:2009zw} and it   corresponds to the eleven dimensional background AdS$_4\times SE_7$ with ${\cal N}=2$ supersymmetry. The $\Schr$ solution has $z=(-4,3)$, where obviously only the value $z=3$ is physical,  and corresponds to the eleven dimensional solution found in \cite{Donos:2009en,Donos:2009xc}, which is discussed from the point of view  of the four dimensional theory in section 4 of  \cite{Gauntlett:2009zw}.

\section{Embeddings Into String/M-theory} \label{sec:embed}
Having established a wide class of supersymmetric solutions in gauged supergravity,
the natural next step is to embed them into string theory or M-theory. The Lifshitz solutions of section \ref{sec:Lifshitz} 
require purely electric or purely magnetic gaugings. One can achieve a purely electric gauging in a simple way by first reducing
 IIB on a Sasaki-Einstein five-manifold ($SE_5$)  \cite{Cassani:2010uw,Gauntlett:2010vu, Liu:2010sa, Skenderis:2010vz} where one obtains $\N=4$ gauged supergravity with two vector-multiplets. Then there is a further truncation \cite{Cassani:2010uw} to an $\N=2$ theory with just the universal hypermultiplet which is gauged electrically under the graviphoton. Dimensional reduction on a circle, with a linear profile for a hyper-scalar, introduces a further electric gauging. Specifically, suppose we take a hyper-scalar
$q$  in five dimensions  and then reduce on the circle 
\bea
ds_5^2&=&ds_4^2+(d\sig+A_1)^2 \, \\
 q&=& k \sig + \tq  \, , 
\eea
where $\tq$ only depends on the co-ordinates of the four-dimensional space-time. It is easy to see that in four dimensions we obtain a kinetic term for $\tq$ of the form
\bea
\cL_4&\sim& (d\tq-k A_1)\w *(d\tq-kA_1) \, ,
\eea
and so $\tq$ has electric charge $k$ under $A_1$. In this way one can obtain a four dimensional
$\N=2$ gauged supergravity theory with cubic prepotential and electric gaugings from IIB on $SE_5\times S^1$.  The $\Lifz$ solutions found in \cite{Donos:2010tu} can probably be understood in this way.

As already discussed in section \ref{schrsection}, certain gaugings of the form \eq{gravel} arise from Sasaki-Einstein reductions of M-theory \cite{Gauntlett:2009zw} and also in the $SU(3)$-invariant sector of the $\N=8$ theory \cite{Bobev:2010ib, Warner:1983vz, Bobev:2009ms}. This makes it clear that our $\Schr$  solutions can be embedded into these theories. It would be interesting to precisely establish which solutions of \cite{Hartnoll:2008rs, Taylor:2008tg, Donos:2009en, Gauntlett:2009zw,Bobev2009a, Bobev2009b,  Donos:2009xc, Donos:2009zf, Donos:2010tu, Donos:2010ax, Jeong:2009aa, O'Colgain:2009yd, Ooguri:2009cv, Colgain:2009wm, Balasubramanian:2010uk, Kraus:2011pf} lie within our class of solutions.

From consistent truncation of type IIA on various nearly-K\"ahler manifolds and cosets \cite{House:2005yc, KashaniPoor:2007tr, Cassani:2009ck}, one can obtain $\N=2$ gauged supergravity with the same scalar manifold we have considered in this work and a rich spectrum of possible electric and magnetic  non-compact gaugings. In the case of purely non compact gaugings we have found no non-relativistic, supersymmetric solutions. It would be  interesting to understand 
if this result holds in general for models with non compact isometries, since these arise naturally in string compactifications.

\vskip 1cm
\noindent {\bf Acknowledgements}
We wish to thank D.~Cassani, G.~Dall'Agata, J.~Gauntlett, A.~Kashani-Poor, S.~Ross, H.~Samtleben, H.~Triendl and A.~Tomasiello for interesting discussions. N.~H~and A.~Z~would like to acknowledge the hospitality of the Galileo Galilei Institute for Theoretical Physics during the course of this project.
The work of N.~H.~is supported by the grant number ANR-07-CEXC-006 of the Agence Nationale de La Recherche. 
M. Petrini is partially supported by the Institut de Physique Th\'eorique, du CEA. A.~Z.~is supported in part by INFN. 

\begin{appendix}
\section{Spinor Conventions}
Our conventions closely follow \cite{Andrianopoli:1996vr, Andrianopoli:1996cm}. We work with in signature $(+---)$. Spinors have the following properties
\bea
\gamma_5 \epsilon_A&=&  \epsilon_A\,, \\
\gamma_5 \epsilon^A&=&- \epsilon^A \,,\\
\epsilon^A&=& (\epsilon_A)^C\,,
\eea
where $\gam_5=-i\gam_0\gam_1\gam_2\gam_3$ and conjugation is defined on a general spinor $\lam$ as
\be
\lambda^C= \gamma_0 C^{-1} \lambda^*\,,
\ee
and
\be
CC^{\dagger}=1, \ \ \ C^2=-1,\ \ \ C^t=-C.
\ee
The gamma matrices satisfy
\bea
\gam_0&=&\gam_0^{\dagger}\,, \\
\gam_i&=& \gam_0 \gam_i^{\dagger}\gam_0.
\eea

\section{Hypermultiplet Scalar Manifold} \label{app:hyper}

Here we summarize various facts about the hypermultiplet scalar manifold $\cM_{\rm Q}$. The eight Killing vectors are given by \cite{Behrndt:2000ph,BrittoPacumio:1999sn}
\be
\begin{array}{ll}
\label{Kv}
k_1=\frac{1}{2i} \blp z_2 \del_{z_1} + z_1 \del_{z_2}-c.c. \brp\, , &k_2=\frac{1}{2} \blp -z_2 \del_{z_1} + z_1 \del_{z_2}+c.c. \brp \, ,   \\ 
k_3=\frac{1}{2i} \blp -z_1 \del_{z_1} + z_2 \del_{z_2}-c.c. \brp\, , &k_4=\frac{1}{2i} \blp z_1 \del_{z_1} + z_2 \del_{z_2}-c.c. \brp \, ,   \\ 
k_5=\frac{1}{2} \blp (-1+z_1^2) \del_{z_1} + z_1 z_2 \del_{z_2}+c.c. \brp\, , &k_6=\frac{i}{2} \blp (1+z_1^2)\del_{z_1} + z_1 z_2\del_{z_2}-c.c. \brp \, ,   \\ 
k_7=\frac{1}{2} \blp -z_1 z_2 \del_{z_1} + (1-z_2^2) \del_{z_2}+c.c. \brp\, , &k_8=\frac{i}{2} \blp z_1 z_2 \del_{z_1} + (1+z_2^2) \del_{z_2}-c.c. \brp \, .
\end{array}
\ee
All these Killing vectors are real, $(k_1,k_2,k_3,k_4)$ generate compact isometries while $(k_5,k_6,k_7,k_8)$ generate non-compact isometries. With the re-definitions
\bea
k_1= -iF_1\,,&& k_2=-iF_2\,, \non \\
k_3= -i F_3 \,,&&k_4=\frac{i}{\sqrt{3}}F_8 \,, \non \\
k_5=F_4 \,,&&k_6=F_5\,, \non \\
k_7= F_6 \,,&&k_8=F_7\, ,\non 
\eea
the commutation relations are $ [F_i,F_j]= if_{ijk} F_k$ with
\bea
&& f_{123}=1,\ \ f_{147}=\frac{1}{2},\ \ f_{156}=-\frac{1}{2},\ \ f_{246}=\frac{1}{2},\\
&& f_{257}=\frac{1}{2},\ \ f_{345}=\half \,,\ \ f_{367}=\half\,, \\
&&f_{458}=\frac{\sqrt{3}}{2}\,, \ \ f_{678}=-\frac{\sqrt{3}}{2}\, .
\eea
Thus we see that $(F_1,F_2,F_3)$ generate $SU(2)$ and $F_8$ generates a commuting $U(1)$.

The Killing prepotentials can be computed from 
\be
\Om^x_{uv}k^u_\Lam =-\nabla_v P^x_{\Lam} \, , 
\ee
where
\be
\Om^x=J^i_{mn}e^m\w e^n,
\ee
$J^i$ are a triplet of complex structures, $e^i$ are frames on $\cM_{\rm Q}$ and $\nabla_v$ is a covariant derivative w.r.t. the $SU(2)$-connection on $\cM_{\rm Q}$. The Killing prepotentials are only well defined up to a local $SU(2)$ transformation. In a particular gauge, the Killing prepotentials  associated to the compact generators are given by (using $r^2 = |\zeta_1|^2+|\zeta_2|^2$)
\bea
P_{1}&=&\frac{1}{r^2\sqrt{1-r^2}} 
\bpm 
\Im (\zeta_1^2- \zeta_2^2)  \\ 
 -\Re  (\zeta_1^2-\zeta_2^2)  \\
\frac{
(r^2-2)\Re(\zeta_2\ol{\zeta}_1)
}{\sqrt{1-r^2}} 
\epm\,, \\
P_{2}&=&\frac{1}{r^2\sqrt{1-r^2}} 
\bpm 
\Re (\zeta_1^2+ \zeta_2^2)  \\ 
 \Im  (\zeta_1^2+ \zeta_2^2)  \\
\frac{
(r^2-2)\Im(\zeta_2\ol{\zeta}_1)
}{\sqrt{1-r^2}} 
\epm\,, \\
P_{3}&=&\frac{1}{r^2\sqrt{1-r^2}} 
\bpm 
2\Im (\zeta_1 \zeta_2)  \\ 
-2 \Re  (\zeta_1 \zeta_2)  \\
\frac{|\zeta_1|^2(2-|\zeta_1|^2)-|\zeta_2|^2(2-|\zeta_2|^2)}{2\sqrt{1-r^2}} 
\epm\,, \\
P_{4}&=&  
-\half\bpm 
0 \\ 
0\\ 
 \frac{r^2}{1-r^2}
\epm  \,, 
\eea
and those associated to the non-compact generators are
\bea
P_{5}=  
\frac{1}{\sqrt{1-r^2}}\bpm 
\Re \zeta_2 \\ 
\Im \zeta_2\\ 
 \frac{\Im \zeta_1}{\sqrt{1-r^2}}
\epm  \,, &&  
P_{6}=  
\frac{1}{\sqrt{1-r^2}}\bpm 
\Im \zeta_2 \\ 
-\Re \zeta_2\\ 
 \frac{\Re \zeta_1}{\sqrt{1-r^2}}
\epm  \,, \\
P_{7}=  
\frac{1}{\sqrt{1-r^2}}\bpm 
\Re \zeta_1 \\ 
\Im \zeta_1\\ 
- \frac{\Im \zeta_2}{\sqrt{1-r^2}}
\epm  \,, &&
P_{8}=  
\frac{1}{\sqrt{1-r^2}}\bpm 
-\Im \zeta_1 \\ 
\Re \zeta_1\\ 
 \frac{\Re \zeta_2}{\sqrt{1-r^2}}
\epm  \,. 
\eea

\end{appendix}

\providecommand{\href}[2]{#2}\begingroup\raggedright\endgroup
\end{document}